\documentclass[10pt,a4paper]{article}
\usepackage[dvips]{graphicx}
\usepackage{amssymb,amsmath}
\usepackage{natbib}
\usepackage{lscape}
\usepackage{subfigure}
\usepackage{rotating}
\usepackage{multirow}
\usepackage[usenames]{color}
\usepackage{setspace}
\usepackage[left=3cm,top=3cm,bottom=3cm]{geometry}
\usepackage{ragged2e}
\begin{document}

\noindent This manuscript version is distributed under the CC-BY-NC-ND (Creative Commons) license.\\

\noindent It has appeared in final form as:\\
Van Albada SJ, Robinson PA. Mean-field modeling of the basal
ganglia-thalamocortical system. I. Firing rates in healthy and parkinsonian states (2009) J
Theor Biol 257: 642--663, DOI: 10.1016/j.jtbi.2008.12.018\\

\newpage

\title{Mean-field modeling of the basal ganglia-thalamocortical system. I. Firing rates in healthy and parkinsonian states}
\author{S. J. van Albada$^{a,b}$ \and P. A. Robinson$^{a,b,c}$}
\maketitle

\begin{center}
$^a$School of Physics, The University of Sydney\\
New South Wales 2006, Australia\\
$^b$The Brain Dynamics Centre, Westmead Millennium Institute\\
Westmead Hospital and Western Clinical School of the University of Sydney\\
Westmead, New South Wales 2145, Australia\\
$^c$Faculty of Medicine, The University of Sydney\\
New South Wales 2006, Australia\\

\end{center}

\begin{abstract}
Parkinsonism leads to various electrophysiological changes in the basal ganglia-thal\-amo\-cort\-ical system (BGTCS), often including elevated discharge rates of the subthalamic nucleus (STN) and the output nuclei, and reduced activity of the globus pallidus external segment (GPe). These rate changes have been explained qualitatively in terms of the direct/indirect pathway model, involving projections of distinct striatal populations to the output nuclei and GPe. Although these populations partly overlap, evidence suggests dopamine depletion differentially affects cortico-striato-pallidal connection strengths to the two pallidal segments. Dopamine loss may also decrease the striatal signal-to-noise ratio, reducing both corticostriatal coupling and striatal firing thresholds. Additionally, nigrostriatal degeneration may cause secondary changes including weakened lateral inhibition in the GPe, and mesocortical dopamine loss may decrease intracortical excitation and especially inhibition. Here a mean-field model of the BGTCS is presented with structure and parameter estimates closely based on physiology and anatomy. Changes in model rates due to the possible effects of dopamine loss listed above are compared with experiment. Our results suggest that a stronger indirect pathway, possibly combined with a weakened direct pathway, is compatible with empirical evidence. However, altered corticostriatal connection strengths are probably not solely responsible for substantially increased STN activity often found. A lower STN firing threshold, weaker intracortical inhibition, and stronger striato-GPe inhibition help explain the relatively large increase in STN rate. Reduced GPe-GPe inhibition and a lower GPe firing threshold can account for the comparatively small decrease in GPe rate frequently observed. Changes in cortex, GPe, and STN help normalize the cortical rate, also in accord with experiments. The model integrates the basal ganglia into a unified framework along with an existing thalamocortical model that already accounts for a wide range of electrophysiological phenomena. A companion paper discusses the dynamics and oscillations of this combined system.
\end{abstract}

\section{Introduction}

The basal ganglia have been studied extensively in connection with a variety of motor and cognitive disorders, including Parkinson's disease (PD), Huntington's disease, and schizophrenia \citep{Bar-Gad2003, Goldman-Rakic1990, Gray1991, Graybiel1990, Haber2004, Swerdlow1987, Walters2007, Waters1988}. In PD, degeneration of dopaminergic neurons in the substantia nigra pars compacta (SNc) leads to changes in tonic and phasic neuronal discharges in the components of the basal ganglia-thalamocortical system (BGTCS). Many studies have provided detailed descriptions of changes in discharge patterns with parkinsonism, as well as suggestions for the pathways and mechanisms by which these patterns arise \citep{Bar-Gad2003, Bergman2002}. An influential proposal is the direct/indirect pathway model of \citet{Albin1989}, which postulates distinct pathways through two populations of striatal neurons expressing either the D1 class or D2 class of dopamine receptor. D1-expressing neurons project monosynaptically to the globus pallidus internal segment (GPi) and the substantia nigra pars reticulata (SNr), giving rise to the \emph{direct pathway}, whereas D2-expressing neurons project polysynaptically to these output nuclei via the globus pallidus external segment (GPe) and the subthalamic nucleus (STN), forming the \emph{indirect pathway}. By enhancing transmission through D2 cells and reducing transmission through D1 cells, degeneration of nigrostriatal dopaminergic neurons would decrease the GPe firing rate in this model, and increase the rates of STN and the output nuclei. This would amplify the inhibitory effect exerted by the output nuclei on the thalamus, leading to parkinsonian symptoms such as akinesia and tremor. The direct/indirect pathway model was later modified to include the so-called \emph{hyperdirect pathway}, upon the realization that the STN receives input directly from the cortex, forming another major input station of the basal ganglia \citep{Nambu2000}. The direct, indirect, and hyperdirect pathways are illustrated in Figure \ref{fig:diagram}.

Physiologically-based mathematical models allow electrophysiological phenomena to be studied not just qualitatively but also in quantitative terms, thus better clarifying the underlying mechanisms. Most computational studies of the basal ganglia consider networks of neurons. \citet{Terman2002} presented a model of the network formed by the STN and the GPe, which displayed either $<$1 Hz or 4--6 Hz oscillations upon dopamine depletion, depending on the network architecture and connection strengths. \citet{Rubin2004} described a neuronal network model that included also the GPi and the thalamus, and illustrated how high-frequency stimulation of the STN may facilitate signal transmission by the thalamus in parkinsonian patients. Specific aspects of basal ganglia function, such as visual attention \citep{Jackson1994} and decision threshold tuning \citep{Lo2006} have also been addressed in computational studies. \citet{Leblois2006} presented a neuronal network model that could account for loss of action selection and predicted the appearance of $\sim$7--10 Hz oscillations in the hyperdirect loop after dopamine depletion. In a detailed model building on earlier work \citep{Gurney2001a, Gurney2001b, Humphries2001}, \citet{Humphries2006} reproduced the enhanced $\sim$1 Hz activity that is observed in the STN and globus pallidus (GP; the rodent homolog of the GPe) of anesthetized rats with nigrostriatal lesions, and gamma-band activity in the healthy state.  

The purpose of this paper is to describe a physiologically plausible mean-field model of the intact BGTCS that can reproduce firing rates characteristic of PD with realistic changes in parameters relative to the non-parkinsonian case. A mean-field model has the advantage over neuronal network models that it can predict large-scale properties of neuronal assemblies and directly assess their dependence on connection strengths between populations. Moreover, mean-field models have comparatively few parameters, and can be implemented for a larger number of populations and connections without leading to an overly complicated set of equations or excessive computational demands. Thus, both numerical and analytical results are more readily obtained. An existing mean-field model has been used successfully to describe thalamocortical oscillations contributing to the electroencephalogram (EEG), yielding predictions of cortical frequency and wavenumber spectra \citep{O'Connor2002, Robinson2001}, coherence and correlations \citep{Robinson2003e}, the electrophysiology of epileptic seizures \citep{Breakspear2006,Robinson2002}, evoked response potentials and steady-state evoked potentials \citep{Kerr2008,Robinson2001b}, and changes with arousal \citep{Robinson2005}. This study provides a first step towards integrating the thalamocortical system and the basal ganglia in a unified framework.

Our model adds to existing models by estimating the strengths of a large number of connections in the BGTCS, and investigating the dependence of average firing rates on these connection strengths. The influences of various projections not present in the classic direct/indirect pathway model are explored, and possible reasons for the prominent hyperactivity of the STN in parkinsonism are discussed. Estimates of parameters and firing rates are based on an extensive review of the experimental literature. Changes in firing rates with nigrostriatal dopaminergic denervation are an aspect of the electrophysiology of the BGTCS that remains to be explained quantitatively, and as such are of key scientific interest. Moreover, steady states form an essential basis for the analysis of dynamics and oscillations, which are the subject of a companion paper [\citet{vanAlbada2009}; henceforth referred to as Paper II].

It is almost impossible for a model of a system as complex as the basal ganglia to incorporate all relevant data, especially as new discoveries are regularly being made. Thus we attempt to distill the main findings from the wealth of available data, while providing a framework that allows for more detailed modeling of basal ganglia structure, activity, and function in future. 

The physiological background of our model is presented in Sec. \ref{sec:phys_background}. Section \ref{sec:model} details the model equations and the possible effects of dopamine depletion. The model is then used to derive firing rates of the BGTCS in the normal and parkinsonian states in Sec. \ref{sec:firing_rate_results}. As mentioned above, these results lay the foundation for the analysis of dynamics and oscillations in Paper II.

\section{Physiological background}\label{sec:phys_background}

This section describes the physiological background of the BGTCS on which our model is based, which allows us to compare the predictions of our model with experimental results. Section \ref{sec:connect} details the main functional connections between the basal ganglia nuclei, its thalamic projection sites, and the cerebral cortex. Section \ref{sec:rates} is devoted to the firing rates of the various components in the normal state and in PD.

\subsection{Functional connectivity of the basal ganglia}\label{sec:connect}

\begin{figure}[ht]
\centering
\includegraphics[width=330pt, height=210pt]{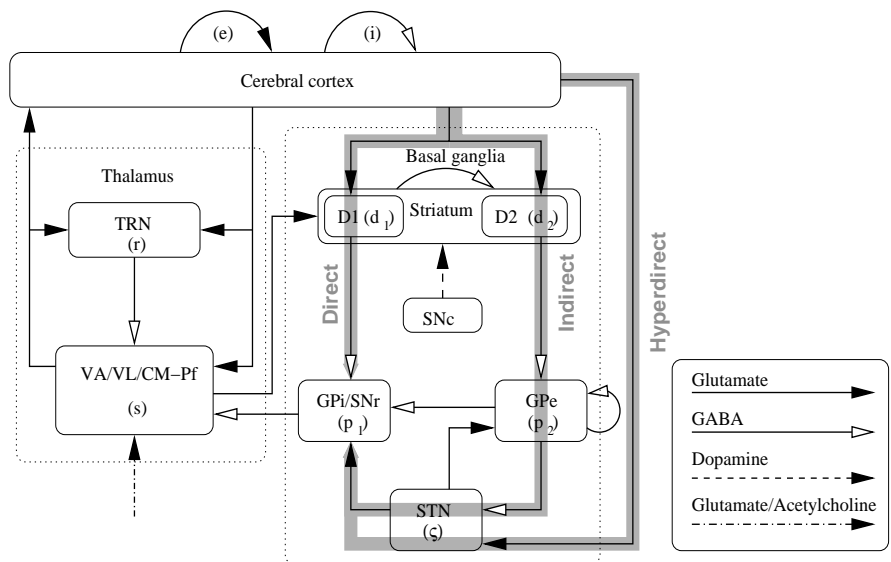} 
\caption{Major functional connections of the BGTCS. External input reaches the thalamus primarily from the brainstem. Filled arrowheads represent excitatory projections, open arrowheads inhibitory ones. Subscripts corresponding to each component are given in parentheses, and gray arrows indicate the direct, indirect, and hyperdirect pathways.}
\label{fig:diagram}
\end{figure}

The main structures comprising the basal ganglia are the striatum (caudate nucleus, putamen, and ventral striatum), pallidum (internal and external segments and ventral pallidum), substantia nigra (pars compacta, pars reticularis, and pars lateralis), and subthalamic nucleus. They are part of a system of pathways, some of which form closed loops, connecting the basal ganglia with the cerebral cortex and thalamus. Information flow through the basal ganglia has been described as following three parallel, mostly separate, pathways (sensorimotor, association, and limbic), which may be further subdivided into somatotopically organized pathways or pathways concerned with different aspects of motor function and cognition \citep{Alexander1986, Alexander1990}. 
The main functional connections of the BGTCS are depicted in Fig. \ref{fig:diagram}.

The SNc and its medial extension, the ventral tegmental area (VTA), send important dopaminergic projections to the striatum \citep{Gerfen1992, Haber2000, Hanley1997}. Excitatory input from the cortex also reaches the basal ganglia mainly at the striatum; sensorimotor inputs terminate more specifically in the putamen, which also receives some associative input \citep{Percheron1984}. The striatum is organized into `patch' and `matrix' compartments, which are distinguished on the basis of biochemical markers and their detailed sources and targets of activity \citep{Gerfen1987}. More than 90\% of striatal neurons are medium spiny neurons \citep{Yelnik1991}, which can be classified both according to their compartmental origin and the class of dopamine receptor they primarily express (D1 or D2). These classifications are partly overlapping: both patch and matrix contain D1 and D2 receptors, although relative receptor densities may differ between compartments \citep{Joyce1988}. Neurons with D1-type receptors coexpress the peptides dynorphin and substance P; D2 cells are enriched in enkephalin \citep{Gerfen1990}. According to the classic direct and indirect pathway model \citep{Albin1989,Alexander1990}, D1 neurons project primarily to the output nuclei GPi and SNr, whereas D2 neurons project primarily to the GPe. Striatal impulses exert an overall excitatory effect on the thalamus and cortex via the direct pathway from the striatum to the output nuclei, and an inhibitory effect via the indirect pathway to the output nuclei via the GPe and the STN (cf. Fig. \ref{fig:diagram}).

In the direct/indirect pathway model, the SNc would mainly facilitate corticostriatal transmission to D1 cells and inhibit transmission to D2 cells, so that dopamine loss would favor the indirect pathway. This simplified view has been called into question by findings that the segregation between D1 and D2 receptors is incomplete \citep{Aizman2000, Inase1997, Surmeier1992, Surmeier1996}, and that neurons expressing both receptor types project to both pallidal segments \citep{Levesque2005,Nadjar2006,Wu2000}. The extent of colocalization of D1 and D2 class receptors reported in the literature ranges from almost none \citep{Hersch1995,LeMoine1995}, to 20--35\% \citep{Inase1997, Lester1993, Meador-Woodruff1991}, about half \citep{Surmeier1996}, or nearly all medium spiny neurons \citep{Aizman2000}. Some of these discrepancies may be explained by a lack of sensitivity of in situ hybridization techniques to low levels of mRNA, which are detected after mRNA amplification \citep{LeMoine1995}, suggesting that even in cells where these receptors occur together, one type usually predominates. Therefore we assume that a significant proportion of striatal neurons expresses a large majority of either D1 or D2 class dopamine receptors.

Despite the collateralization of striatofugal axons, many studies have also shown that projections in the direct and indirect pathways can be at least partly distinguished. In a rat model of PD, striatopallidal neurons show increased expression of mRNA encoding D2 receptors and enkephalin, whereas striatonigral neurons show a reduction in mRNA for D1 receptors and substance P \citep{Gerfen1990}. In Huntington's disease the striatal projection to GPe is more vulnerable than that to GPi \citep{Deng2004,Reiner1988,Walker2007}. In a study of mouse brain slices, \citet{Day2006} found that lack of dopamine causes a profound loss of dendritic spines on striatopallidal neurons but not on striatonigral neurons. Thus, we assume a partial segregation of the projections to the output nuclei and the GPe. However, our model provides an overarching framework in which both possibilities (segregation or overlap) can be incorporated, and differences between these possibilities can be explored.

Besides medium spiny neurons, the striatum contains various types of interneurons, including cholinergic tonically active neurons that make up about 1--5\% of the striatum \citep{Aosaki1995, Kawaguchi1995, Kimura1984}, and GABAergic inhibitory interneurons that make up only a small percentage of the striatal population but have strong effects \citep{Bolam2000, Koos1999}. In addition, medium spiny neurons have local axon collaterals through which GABA exerts a depolarizing effect at rest, but a hyperpolarizing effect near spike threshold \citep{Plenz2003,Taverna2004}. Thus, lateral connections between medium spiny neurons will moderate the striatal firing rate with strong cortical inputs.

In primates the SNr and GPi are part of separate circuits, with different target areas and sources of activity \citep{Ilinsky}. The SNr receives input mostly from the caudate nucleus, which relays associative information from the prefrontal cortex as well as inputs from the frontal eye fields. It sends GABAergic projections mainly to the magnocellular part of the ventral anterior nucleus (VAmc) of the thalamus and is involved in the control of eye movements \citep{Parent1995}. The GPi, on the other hand, receives input mainly from the premotor and primary motor cortices via the putamen, and relays this mainly to the ventrolateral thalamic nucleus (VL) \citep{Haber2004}. Despite these differences, GPi and SNr are often modeled as a single structure due to their closely related inputs and outputs, as well as similarities in cytology and function \citep{Alexander1990, Bar-Gad2003}. Since electrophysiological studies of the remaining basal ganglia nuclei often do not distinguish between associative and sensorimotor territories, it is difficult in practice to differentiate between the inputs to GPi and SNr. We therefore model these nuclei as a single combined structure, although the response to dopaminergic cell loss is more pronounced in the GPi \citep{Mitchell1986,Wichmann1999}. 

Apart from the ventral anterior nucleus (VA) and VL, target sites of the basal ganglia output nuclei have been identified in the centromedian-parafascicular complex (CM-Pf) \citep{Kim1976, Parent2001}. Neurons in VA, VL, and CM-Pf send axons back mainly to the matrix compartment of the striatum \citep{Carpenter, Gonzalo2002, McFarland2000, Parent2, Ragsdale1988, Sadikot1992}. Studies suggest that the influence of these projections is excitatory \citep{Haber2004, Sadikot1992}. 

The GPe sends an important inhibitory projection to the STN, which in turn excites both the GPe and the output nuclei \citep{Hamada1992b, Kita1983b,Parent1995,Shink1996}.
However, the pattern of connections between GPe, GPi, and STN is complicated by a direct projection from approximately a third of GPe neurons forming synapses on GPi cell bodies or proximal dendrites \citep{Hazrati1990, Sato2000b, Shink1995, Smith1994}. These projections derive from axons also branching to STN and sometimes SNr \citep{Sato2000b}. Besides its substantial innervation by striatum and STN, the GPe is extensively connected via local axon collaterals, which may exert a strong inhibitory influence since they terminate on cell bodies and proximal dendrites \citep{Kita1994, Nambu1997, Ogura2000}.

As discussed in the Introduction, the STN forms an additional input station of the basal ganglia. The cortico-STN projection originates in the primary motor cortex (M1) and somatosensory and premotor cortices, including the supplementary motor area (SMA) \citep{Afsharpour1985, Nambu1996, Nambu1997b, Nambu2000, Parent1995}. Because the STN influences the thalamus mainly via direct projections to the GABAergic output nuclei, the overall effect of this pathway on thalamic targets is inhibitory.

Connections within and between the thalamus and cortex complete the basal ganglia-thalam- ocortical system. These connections follow a previous model of brain electrical activity involving only the thalamus and cortex \citep{Rennie1999,Robinson1997,Robinson2001, Robinson2003b, Robinson2005}. The thalamic reticular nucleus (TRN) exerts a powerful inhibitory effect over the relay nuclei, from which it receives excitatory input. Both TRN and the relay nuclei are densely innervated by glutamatergic cortical neurons. Within the cortex our model includes excitatory corticocortical and inhibitory local circuit neurons. Finally, sensory stimuli reaching the thalamus mainly from the brainstem are modeled as external input. 

\subsection{Data on firing rates in normal and parkinsonian states}\label{sec:rates}

This section provides a summary of the mean firing rates of the basal ganglia nuclei and their thalamic and cortical targets in the normal and parkinsonian states, for comparison with modeling results in Sec. \ref{sec:firing_rate_results}.

Some studies of firing rates and patterns of basal ganglia are performed during stereotaxic surgery for PD. However, most studies use one of two well-known animal models of parkinsonism. In monkeys, symptoms most closely resembling human parkinsonism are obtained by lesioning nigrostriatal neurons using \mbox{1-methyl-4-phenyl-1,2,3,6-tetrahydropyridine} (MPTP) \citep{DeLong1990}. Depending on the species, this can lead to akinesia, bradykinesia, and/or resting tremor at frequencies of \mbox{4--8 Hz}. Another widely used paradigm is the 6-hydroxydopamine (6-OHDA) rodent model of PD \citep{Ungerstedt1968}. 

The average discharge rate of neurons in the primary motor and somatosensory cortices of monkeys is about 5--20 s$^{-1}$ depending on the level of activity \citep{Wannier1991}. Some studies found the cortical rate to be unchanged with MPTP- or 6-OHDA-induced parkinsonism \citep{Dejean2008, Goldberg2002}, and parkinsonian symptoms have instead been linked with abnormal temporal organization of motor cortical activity \citep{Brown2000, Goldberg2002, Salenius2002}. On the other hand, fMRI studies in PD patients have found impaired activation of cortical areas normally co-activated with the striatum \citep{Monchi2004,Monchi2007}, and a PET study showed that blood flow in the SMA of PD patients during performance of a motor task is reduced relative to healthy subjects \citep{Jenkins1992}.

Most striatal medium spiny neurons fire spontaneously at a low rate of 0.5--2 s$^{-1}$ \citep{DeLong1983,Haber2004, Kimura1996}. However, \citet{Kiyatkin1999} reported a highly skewed distribution of striatal firing rates in awake rats, with a small fraction of fast-spiking neurons taking the average rate up to $\sim$6 s$^{-1}$. This matches the rate in monkeys recorded by \citet{Goldberg2002}. Several studies in 6-OHDA-lesioned rats have revealed elevated activity in striatal neurons with respect to healthy rats \citep{Chen2001,Kish1999,Tseng2001,Walters2007}, which may be due to a large activity increase in striatopallidal neurons and a smaller activity decrease in striatonigral neurons \citep{Mallet2006}. A relatively high average firing rate of $\sim$10 s$^{-1}$ was also recorded in the putamen of PD patients \citep{Magnin2000}. On the other hand, one study reported the discharge rate of caudate neurons to be decreased from $\sim$6 s$^{-1}$ to $\sim$4 s$^{-1}$ with MPTP lesion \citep{Yoshida1991}, while another study found no change \citep{Goldberg2002}. 

The normal GPi firing rate in primates is in the range 60--90 s$^{-1}$, with an increase of about 10--20 s$^{-1}$ after MPTP treatment in monkeys \citep{Filion1991,Heimer2002,Yoshida1991}. 
In line with these results, normalization of dopamine levels by application of the dopamine agonist apomorphine decreases the average firing rate of GPi neurons in PD patients \citep{Merello1999}. 
On the other hand, some studies found no significant change in average firing rates of GPi \citep{Bergman1994, Wichmann1999} or its rodent homolog, the entopeduncular nucleus (EP) \citep{Robledo1991}. In one study the mean discharge frequency of GPi neurons in PD patients was only $\sim$59 s$^{-1}$ \citep{Sterio1994}, and another study \citep{Hutchison1994} found the average firing rate of GPi neurons in PD patients to be 67 s$^{-1}$ and no different from the average rate reported for normal monkeys. However, the posteroventral portion of the GPi showed increased activity (mean rate 82 s$^{-1}$). This suggests that dopamine loss increases the firing rate in the sensorimotor portion of the GPi, while other parts are relatively unaffected. \citet{Magnin2000} similarly differentiated between the GPi as a whole, which had a firing rate of 91 s$^{-1}$ in PD patients, and the internal part of the GPi, which discharged at 114 s$^{-1}$.

SNr neurons in normal monkeys discharge at a mean rate comparable to that of GPi neurons (50--70 s$^{-1}$) \citep{DeLong1983, Schultz1986}. However, the SNr is less affected by dopamine lesions than other nuclei.  For instance, the mean firing rate of SNr neurons in PD patients is $\sim$71 s$^{-1}$ \citep{Hutchison1998}, very close to that in normal monkeys. In a study by \citet{Walters2007}, SNr neurons of 6-OHDA-lesioned rats displayed a non-significant decrease in average firing rate 7--10 days post-lesion. \citet{MacLeod} measured a short-term decrease ($<10$ days postlesion) in the average firing rate of SNr neurons upon treatment with 6-OHDA, but the firing rate had normalized after a period of $>6$ months. On the other hand, \citet{Benazzouz2000} and \citet{Burbaud1995} recorded significantly elevated discharge rates in the SNr of rats several weeks after lesions of the SNc. Alterations in SNr firing rates and patterns reported by \citet{Wichmann1999} were less pronounced than those in the GPi, and no significant change in SNr firing rate was observed.

GPe neurons can be divided into two main categories based on firing characteristics \citep{DeLong1971, Filion1991, Sterio1994}: about 85\% of GPe neurons display high-frequency bursts of activity interspersed with long intervals of silence lasting up to several seconds. These neurons have a mean firing rate of $\sim$55 s$^{-1}$. The remaining 15\% are slowly discharging neurons with occasional bursts and an average rate of $\sim$10 s$^{-1}$. Conflicting reports exist concerning changes in GPe rate, some studies finding a decrease of about  10--20 s$^{-1}$ with nigrostriatal lesions \citep{Boraud1998, Filion1991, Heimer2002, Pan1988}, whereas others detected no significant change \citep{Goldberg2002, Hutchison1994, Magill2001, Walters2007}. Average GPe firing rates of 40--60 s$^{-1}$ have been reported in patients with medication-resistant PD \citep{Hutchison1994, Magnin2000, Sterio1994}.

STN cells in monkeys display spontaneous tonic activity, firing at approximately 20--30 s$^{-1}$, often in pairs or triplets of spikes \citep{DeLong1985, Georgopoulos1983}. Dopaminergic lesion has been reported to increase this rate by $\sim$7 s$^{-1}$ \citep{Bergman1994}. In agreement with these findings, STN neurons of PD patients have relatively high discharge rates of 37--43 s$^{-1}$ \citep{Benazzouz2002,Hutchison1998}, while \citet{Levy2000} measured a higher median firing rate in STN cells that displayed tremor-related activity (53 s$^{-1}$) than in non-tremor-related cells (43 s$^{-1}$) of PD patients. Increases of 4--6 s$^{-1}$ in mean STN discharge rate have been observed in 6-OHDA-treated rats \citep{Kreiss1997, Walters2007}, although some studies found no change or even a reduction in firing rate up to 4 weeks postlesion \citep{Hollerman1992, Ni2001}.

Firing in pallidal-receiving areas of the thalamus was found to be 7--8 s$^{-1}$ in PD patients compared with 18--19 s$^{-1}$ in patients with essential tremor or pain \citep{Molnar2005}. Since the basal ganglia are not thought to be involved in the pathophysiology of pain or essential tremor, a rate of 18--19 s$^{-1}$ probably represents normal thalamic activity, suggesting that activity of pallidal-receiving thalamic areas is reduced in PD. A significant decrease in thalamic activity was found in MPTP-treated cats \citep{Schneider1996} but not monkeys \citep{Pessiglione2005}. However, metabolic studies in 6-OHDA-treated rats and MPTP-treated monkeys strongly point to hypoactivity of basal ganglia-receiving areas of the thalamus in parkinsonism \citep{Gnanalingham1995,Palombo1988,Rolland2007}.

The TRN has an average firing rate of about 20--30 s$^{-1}$ in awake cats, which correlates positively with the level of arousal and hence with the activity of the relay nuclei \citep{Steriade1986}. \citet{Raeva1987} measured the activity of TRN neurons during stereotaxic surgery on subjects with dyskinesia, most of whom were parkinsonian. They found three types of cells with different discharge patterns, for which the overall mean firing rate was about 10 s$^{-1}$. Although information on changes in TRN activity with dopamine loss is limited, this may be taken as indirect evidence that the TRN is hypoactive in PD.

Average firing rates of the components of the BGTCS in the normal state and changes with parkinsonism are summarized in Table~1. We do not report the control rates from many of the studies in rats, because they were performed on animals under general anesthesia, which leads to significantly lower firing rates than the freely moving condition \citep{Benazzouz2000, Kreiss1997, Pan1988, Rohlfs1997}. Human control data are not available for most nuclei, since stereotaxic surgery is only performed in clinical cases.

\begin{table}[htp]
\centering
\begin{minipage}{\textwidth} 
\centering
\begin{tabular}{p{1.3cm}p{1.2cm}p{1.6cm}p{1.2cm}p{1.2cm}p{1.1cm}p{1.8cm}p{2.2cm}}
\hline
Location & \multicolumn{4}{c}{Normal rate (s$^{-1}$)} & PD \mbox{patients} & \multirow{2}{1.8cm}{MPTP-treated monkeys} & 6-OHDA-treated rats \\ 
& \emph{Humans} & \emph{Monkeys} & \emph{Cats} & \emph{Rats} & & & \\
\hline
Cortex & & 5--20$^{ab}$ & & 2--5$^c$ & $\downarrow^{def}$ & ---$^a$ & ---$^c$ \\
Striatum & & 4--7$^{ag}$ & & 1--7$^{ch}$ & $\uparrow^i$ & $\downarrow^g$, ---$^a$ & $\uparrow^{jklm}$ \\
GPi & & 60--90$^{nopq}$& & 15--20$^{c}$ & $\uparrow$/---$^r$  & $\uparrow^{gpst}$, ---$^{uv}$ & ---$^w$ \\
SNr & & 50--70$^{xy}$ & & & ---$^z$  & ---$^w$ & $\uparrow^{AB}$, $\downarrow$/---$^{mC}$ \\
GPe & & 40--70$^{noqD}$ & & 35--45$^E$ & ---$^r$ & $\downarrow^{qst}$, ---$^a$ & $\downarrow^E$, ---$^m$ \\
STN & & 20--30$^{oF}$ & & 8--11$^G$ & $\uparrow^{zEH}$ & $\uparrow^u$ & $\uparrow^{mG}$, ---$^I$, $\downarrow^J$ \\
Relay nuclei & 10--20$^K$ & & & & $\downarrow^K$ & $\downarrow^{LMN}$,---$^O$& $\downarrow^N$\\
TRN & & & 20--30$^P$ & & $\downarrow^{Q}$ & & \\
\hline
\end{tabular}
\end{minipage}
\caption{Average firing rates of the BGTCS in the healthy state, and changes with respect to this state in PD patients (compared with normal rates in monkeys, humans, or cats, as available), in MPTP-treated monkeys, and in rats treated with 6-OHDA. In the rows for GPi and GPe, changes in firing rates in rats refer to the rodent homologs of these structures, the entopeduncular nucleus (EP) and the globus pallidus (GP). Changes reflect averages; the reaction to loss of dopamine may differ in individual cases. $\uparrow$, elevated; ---, no change; $\downarrow$, reduced. References: $a$, Goldberg et al. (2002); $b$, Wannier et al. (1991); $c$ Dejean et al. (2008); $d$, Jenkins et al. (1992); $e$ Monchi et al. (2004); $f$ Monchi et al. (2007); $g$, Yoshida (1991); $h$, Kiyatkin and Rebec (1999); $i$, Magnin et al. (2000); $j$ Chen et al. (2001); $k$ Kish et al. (1999); $l$, Tseng et al. (2001); $m$, Walters et al. (2007); $n$, DeLong (1971); $o$, Georgopoulos et al. (1983); $p$, Kimura et al. (1996); $q$, Heimer et al. (2002); $r$, Hutchison et al. (1994); $s$, Filion and Tremblay (1991), $t$, Boraud et al. (1998); $u$, Bergman et al. (1994); $v$, Wichmann et al. (1999); $w$, Robledo and Feger (1991); $x$, DeLong et al. (1983); $y$, Schultz (1986); $z$, Hutchison et al. (1998); $A$, Benazzouz et al. (2000); $B$, Burbaud et al. (1995); $C$, MacLeod et al. (1990); $D$, DeLong et al. (1985); $E$, Pan and Walters (1988); $F$, Benazzouz et al. (2002); $G$, Kreiss et al. (1997); $H$, Levy et al. (2000); $I$, Hollerman and Grace (1992); $J$, Ni et al. (2001a); $K$, Molnar et al. (2005); $L$, Palombo et al. (1988); $M$, Gnanalingham et al. (1995); $N$, Rolland et al. (2007); $O$, Pessiglione et al. (2005); $P$, Steriade et al. (1986); $Q$, Raeva and Lukashev (1987).}
\end{table}

\section{Model formulation and preliminary analysis}\label{sec:model}

In order to arrive at a tractable model of the dynamics governing the system in Fig. 1, we use a mean-field formulation, in which neuronal properties are spatially averaged. The dynamics are then governed by a set of equations relating the average firing rates of populations of neurons to changes in cell-body potential, which are in turn triggered by average rates of incoming pulses. This approach is based on earlier work on a model of the electrophysiology of the corticothalamic system \citep{Rennie1999, Robinson1997, Robinson2001, Robinson2003b, Robinson2005}. Section \ref{sec:equations} details the basic equations of the model. Parameter values for healthy adults in the alert, eyes-open state are estimated in Sec. \ref{sec:parameters}, and used to evaluate fixed points in Sec. \ref{sec:steady_states}.
In Sec. \ref{sec:depletion} we review possible ways of modeling dopamine depletion. 

\subsection{Basic equations}\label{sec:equations}

The first component of the model is the description of the average response of populations of neurons to changes in cell-body potential. The mean firing rate $Q_a(V_a)$ of each population $a$ is taken to be the maximum attainable firing rate $Q_a^{\mathrm{max}}$ times the proportion of neurons with a membrane potential $V_a$ above the threshold potential $x$. Equivalently, the response of each neuron can be represented by a Heaviside step function $H(V_a-x)$ multiplied by $Q_a^{\mathrm{max}}$, and the population rate is given by the integral of this response times the distribution $p(x)$ of firing thresholds,
\begin{equation}
Q_a(V_a) = Q_a^{\mathrm{max}} \int_{-\infty}^{\infty}  H(V_a-x)p(x) dx,
\end{equation}
yielding the cumulative distribution function, which for a Gaussian distribution is the error function. However, the exact distribution of firing thresholds is not known, allowing us to work with the closely similar sigmoidal function
\begin{equation}\label{eq:1}
Q_a(\mathbf{r},t) \equiv S_a[V_a(t)] = \frac{Q_a^{\mathrm{max}}}{1+\mathrm{exp}[-(V_a(t) - \theta_a)/\sigma']},
\end{equation}
for analytical convenience. Here, $\theta_a$ is the mean threshold potential of the population considered. Fitting (\ref{eq:1}) to the error function we find that $\sigma'$ is $\sqrt{3}/\pi$ times the standard deviation of the Gaussian distribution of firing thresholds \citep{Wright1995}. In the absence of detailed information on the standard deviations of firing thresholds in the basal ganglia, we set $\sigma'$ equal for all populations. The function (\ref{eq:1}) increases smoothly from $0$ to $Q_a^{\mathrm{max}}$ as $V_a$ runs from $-\infty$ to $\infty$.

The change in the mean cell-body potential due to afferent activity depends on the mean number of synapses $N_{ab}$ from afferent axons of type $b$ per receiving neuron of type $a$, and the typical time-integrated change $s_{ab}$ in cell-body potential per incoming pulse. Defining $\nu_{ab} = N_{ab}s_{ab}$, the change in the mean cell-body potential of type $a$ neurons is thus modeled as \citep{Robinson2004}
\begin{align}\label{eq:2}
D_{\alpha\beta}(t)V_a(t) = \sum_b \nu_{ab}\phi_b(t-\tau_{ab}),\\
D_{\alpha\beta}(t) = \frac{1}{\alpha\beta}\frac{d^2}{dt^2}+\left(\frac{1}{\alpha} + \frac{1}{\beta}\right)\frac{d}{dt} + 1.
\end{align}
Here, $\phi_b(t-\tau_{ab})$ is the incoming pulse rate, $\tau_{ab}$ represents the axonal time delay for signals traveling from type $b$ to type $a$ neurons, and $\alpha$ and $\beta$ are the decay and rise rates of the cell-body potential (we assume $\alpha < \beta$ without loss of generality). The differential operator $D_{\alpha\beta}(t)$ is a physiologically realistic representation of dendritic and synaptic integration of incoming signals \citep{Rennie2000, Robinson1997}. The synapses and dendrites attenuate high-frequency activity due to differential delays for signals passing through them, forming an effective low-pass filter with cut-off frequency intermediate between $\alpha$ and $\beta$. 
%
%
%
In general, $\alpha$ and $\beta$ can depend on both the sending and receiving neurons, but in order to restrict the number of parameters we take these to be equal for all populations, especially as the values of $\alpha$ and $\beta$ are not relevant to steady states, which are the main focus of the current paper. Future work may use different rise and decay rates for different populations, as was done for instance by \citet{Rennie2000}. 

In a number of previous works, neuronal activity was modeled as spreading along the cortex in a wavelike fashion \citep{Bressloff2001,Bressloff2003,Jirsa1996,Jirsa1997,Nunez1995},
based on consistent experimental observations of such waves of activity upon localized cortical stimulation \citep{Burns1951,Chervin1988,Golomb1997,Lopes1978,Nunez1974,Prechtl1997,Rubino2006,Schiff2007,Wu1999,Xu2007}. Estimates of characteristic axonal ranges and propagation speeds suggest that such waves are significantly damped on the scale of the human cortex \citep{Robinson2001b,Robinson2004,Wright1995}. A damped-wave equation was derived in \citet{Robinson1997} using a range distribution of corticocortical fibers that decayed exponentially at large distances. Ignoring the spatial derivative, which is not relevant in the present context, this equation simplifies to
%
\begin{equation}\label{eq:3}
\frac{1}{\gamma_a^2}\left[\frac{\partial^2}{\partial t^2}+ 2\gamma_a\frac{\partial}{\partial t}+\gamma_a^2 \right]\phi_a(t) = Q_a(t),
\end{equation}
%
%
where $\gamma_a = v_a/r_a$ is the damping rate, consisting of the average axonal transmission speed $v_a\simeq 5-10$ m s$^{-1}$ and the characteristic axonal range $r_a$. In practice, most types of axons are short enough to justify setting $\gamma_a = \infty$, which has been termed the local activity approximation \citep{Robinson2004}. We therefore take only $\gamma_e$, the damping rate associated with cortical pyramidal cells, to yield significant propagation effects. This turns all wave equations except the cortical one into delayed one-to-one mappings. The temporal dependence in Eq. (\ref{eq:3}) is given for completeness, although we are interested in steady states.
 
Besides cortical excitatory cells, our model includes eight neuronal populations: cortical inhibitory ($i$), striatal cells projecting to the output nuclei ($d_1$), striatal cells projecting to GPe ($d_2$), GPi/SNr ($p_1$), GPe ($p_2$), STN ($\varsigma$), thalamic relay nuclei ($s$), and TRN ($r$). The subscript $s$ for the relay nuclei follows the convention of earlier work, in which it referred only to specific relay nuclei, although here we also consider the diffusely projecting CM-Pf complex. Input from the brainstem to the thalamus is denoted by a subscript $n$. For simplicity, we consider inhibition within the striatum of D1 to itself and D2 to itself, but not between D1 and D2.

\subsection{Parameter values}\label{sec:parameters}

Before proceeding to the analysis of fixed points, we here discuss how parameter values were chosen based on known physiology. Among the parameters in our model that can be relatively well measured experimentally are axonal conduction times and maximum firing rates of neuronal populations. The relative strengths of the various connections can be estimated based on experimentally determined densities of projections, types of neurotransmitters, and locations of synapses. Besides, plausible parameters should yield realistic steady-state firing rates both before and after changes that would be expected with loss of dopamine. In this section we constrain parameter values for our model using evidence from a range of studies, leading to the nominal values given in Table~2. Note that the results of the present study are independent of axonal or dendritic delays, but these values are listed for completeness and use in Paper II.

\begin{table}[htp]
\centering
\begin{minipage}{\textwidth} 
\centering
\begin{tabular}{p{2.7cm} p{1.9cm} p{0.8cm} p{0.7cm} p{7.3cm}}
\hline
\multicolumn{5}{c}{\textbf{Table~2. Model parameters for healthy adults in the alert, eyes-open state}}\\
\hline
Quantity & Symbol & Value & Unit & References \\ 
\hline
Corticocortical axonal range & $r_e$ & 80 & mm & Nunez, 1995; O'Connor et al. (2002), Robinson (2003) \\
Cortical damping rate & $\gamma_e$ & 125 & s$^{-1}$ & Robinson et al. (2004), Rowe et al. (2004) \\
Synaptodendritic & & & & \\
\ \ \ Decay rate & $\alpha$ & 160 & s$^{-1}$ & Destexhe and Sejnowski (2001), Hestrin et al. (1990) \\
\ \ \ Rise rate & $\beta$ & 640 & s$^{-1}$ & Destexhe and Sejnowski (2001), Hestrin et al. (1990) \\
Axonal delay & & & & \\
\ \ \ $es, is$ & $\tau_{es}$, $\tau_{is}$ & 35 & ms & Roberts and Robinson (2008), Robinson et al. (2004), Rowe et al. (2004) \\
\ \ \ $d_1e, d_2e$ & $\tau_{d_1e}$, $\tau_{d_2e}$ & 2 & ms & Kimura et al. (1996), Nambu et al. (2000) \\
\ \ \ $d_1s, d_2s$ & $\tau_{d_1s}$, $\tau_{d_2s}$ & 2 & ms & Clugnet et al. (1990) \\
\ \ \ $p_1d_1$ & $\tau_{p_1d_1}$ & 1 & ms & Kimura et al. (1996), Nambu et al. (2000)\\
\ \ \ $p_1p_2$ & $\tau_{p_1p_2}$ & 1 & ms & Kita (2001) \\
\ \ \ $p_1\varsigma$ & $\tau_{p_1\varsigma}$ & 1 & ms & Nambu et al. (2000) \\
\ \ \ $p_2d_2$ & $\tau_{p_2d_2}$ & 1 & ms & Kimura et al. (1996), Nambu et al. (2000)\\
\ \ \ $p_2\varsigma$ & $\tau_{p_2\varsigma}$ & 1 & ms & Nambu et al. (2000) \\
\ \ \ $\varsigma e$ & $\tau_{\varsigma e}$ & 1 & ms & Bar-Gad et al. (2003), Maurice et al. (1998), Nambu et al. (2000) \\
\ \ \ $\varsigma p_2$ & $\tau_{\varsigma p_2}$ & 1 & ms & Nambu et al. (2000) \\
\ \ \ $se$, $re$ & $\tau_{se}$, $\tau_{re}$ & 50 & ms & Roberts and Robinson (2008), Robinson et al. (2004), Rowe et al. (2004)\\
\ \ \ $sp_1$ & $\tau_{sp_1}$ & 3 & ms & Anderson and Turner (1991), Uno et al. (1978) \\
\ \ \ $sr, rs$ & $\tau_{sr}$, $\tau_{rs}$ & 2 & ms & Voloshin and Prokopenko (1978) \\
Maximum firing rate & & & & \\
\ \ \ Cortex & $Q_e^{\mathrm{max}}$, $Q_i^{\mathrm{max}}$  & 300 & s$^{-1}$ & McCormick et al. (1985), Steriade et al. (1998) \\
\ \ \ Striatum & $Q_{d_1}^{\mathrm{max}}$, $Q_{d_2}^{\mathrm{max}}$ & 65 & s$^{-1}$ & Kiyatkin and Rebec (1999) \\
\ \ \ GPi/SNr & $Q_{p_1}^{\mathrm{max}}$ & 250 & s$^{-1}$ & Hashimoto et al. (2003), Nakanishi et al. (1987)\\
\ \ \ GPe & $Q_{p_2}^{\mathrm{max}}$ & 300 & s$^{-1}$ & Cooper and Stanford (2000), Nambu and Llin{\'a}s (1997a)\\
\ \ \ STN & $Q_{\varsigma}^{\mathrm{max}}$ & 500 & s$^{-1}$ & Kita et al. (1983), Nakanishi et al. (1987)\\
\ \ \ Relay nuclei & $Q_s^{\mathrm{max}}$ & 300 & s$^{-1}$ & Destexhe and Sejnowski (2003)\\
\ \ \ TRN & $Q_r^{\mathrm{max}}$ & 500 & s$^{-1}$ & Raeva and Lukashev (1987) \\
Firing threshold & & & & \\
\ \ \ Cortex & $\theta_e$, $\theta_i$ & 14 & mV & \multirow{7}{*}{\begin{sideways} $\underbrace{~~~~~~~~~~~~~~~~~~~~~~}$ \end{sideways}} \\
\ \ \ Striatum & $\theta_{d_1}$, $\theta_{d_2}$ & 19 & mV & \\
\ \ \ GPi/SNr & $\theta_{p_1}$ & 10 & mV & \\
\ \ \ GPe & $\theta_{p_2}$ & 9 & mV & ~~~~From numerical exploration (see caption) \\
\ \ \ STN & $\theta_{\varsigma}$ & 10 & mV & \\
\ \ \ Relay nuclei & $\theta_s$ & 13 & mV & \\
\ \ \ TRN & $\theta_r$ & 13 & mV & \\
\hline
\end{tabular}
\end{minipage}
\end{table}

Conduction delays between neuronal populations can be estimated using antidromic or sometimes orthodromic activation, spike-triggered averaging, or cross-correlation analysis \citep{Nowak1997}. The results of such studies always include axonal propagation times, but may also include dendritic and synaptic delays and neuronal integration times of the sending and/or receiving populations, depending on the method used. It is important to take this into account, since dendritic and synaptic latencies and integration times may be as long or longer than axonal propagation times \citep{Nowak1997}. In addition, care is needed to determine average or characteristic delays rather than the shortest possible ones, since the former are more relevant to ongoing oscillations.

Axonal delays measured using spike timing and correlation in mice, rabbits, cats, and monkeys range from 0.1 to 5 ms for the thalamocortical projection, and 1 to 30 ms for the corticothalamic projection, with longer delays expected after scaling these values to human brain size \citep{Roberts2008}. As argued in that paper, ongoing corticothalamic oscillations depend on a weighted average of conduction velocities of many fibers, further increasing the latency with respect to the values found in most experimental studies, which typically select the shortest ones. In accordance with model fits to absence seizure dynamics \citep{Roberts2008} we split the axonal propagation time for a full loop into a thalamocortical axonal delay of 35 ms and a somewhat longer corticothalamic delay of 50 ms. The hypothesis that the alpha rhythm of the EEG is caused by a resonance in a corticothalamic loop with an axonal delay of $\sim$85 ms has led to excellent agreement with data on a range of electrophysiological phenomena \citep{O'Connor2002, Robinson2001, Robinson2003e, Breakspear2006, Kerr2008}. 

The corticosubthalamic pathway has been shown in some studies to act faster than transmission via the direct or indirect pathways \citep{Bar-Gad2003}. Total cortico-striato-pallidal and cortico-subthalamo-pallidal delays can be inferred from the pattern of GPe responses to cortical activation in healthy monkeys, consisting of an excitation after 8--11 ms, inhibition after 15--19 ms, and a second excitation after 26--32 ms due to disinhibition of the STN \citep{Kita2004,Nambu2000,Yoshida1993}. Since we assume synaptic and dendritic integration times of the order of 6--8 ms (time-to-peak) \citep{Destexhe2001,Hestrin1990}, this implies that the corticosubthalamic and STN-GPe axonal delays are only about 1 ms each. The reciprocal delay from GPe to STN is also $\lesssim$ 1 ms \citep{Nambu2000}. A corticostriatal axonal delay of 2 ms and striatopallidal delay of 1 ms lead to onset of GPe excitation after about 8 ms, inhibition after 19 ms, and a second excitation after 28 ms, in accord with the above studies. The latency of postsynaptic responses in EP neurons to GP stimulation is around 3--6 ms in the rat brain slice preparation \citep{Kita2001}, suggesting an axonal latency of $\lesssim$ 1 ms. \citet{Anderson1991} measured a reduction in the activity of pallidal-receiving thalamic neurons of awake monkeys generally $<$4 ms after stimulation of the GPi, although some cells displayed a longer latency. In anesthetized rats, the average latency of spikes in the striatum evoked by stimulation of the medial geniculate body of the thalamus was $\sim$4 ms \citep{Clugnet1990}. We can roughly translate this into an axonal delay of $\sim$2 ms in humans by noting that the value of 4 ms includes striatal integration times, but on the other hand, delays are likely to be slightly longer in humans than in rats (compare for instance the corticosubthalamic axonal plus dendritic and synaptic delay of $\sim$4.5 ms in rats, \citeauthor{Maurice1998}, \citeyear{Maurice1998} with the $\sim$6 ms delay in monkeys, \citeauthor{Nambu2000}, \citeyear{Nambu2000}). \citet{Voloshin1978} measured orthodromic and antidromic response latencies of TRN neurons to stimulation of VL in cats. Interpretation of the data is complicated by the possible contribution of polysynaptic pathways, but 2 ms seems to be an adequate approximation to the monosynaptic axonal delay.

A number of studies have provided estimates of the resting membrane potentials, firing thresholds, and maximum firing rates of neurons in the basal ganglia. Most of the data come from studies of rodents, but these provide the closest possible approximation to human values, which are often not readily available. Since the biophysics determining these quantities is likely to be very similar across species, we assume that it is reasonable to use data from rodents.

The maximum firing rate of cortical regular spiking neurons is of order 250 s$^{-1}$ \citep{McCormick1985}, whereas a class of corticothalamic fast rhythmic bursting neurons can fire up to $\sim$400 s$^{-1}$ in either burst or tonic mode. These populations are not strictly distinct, since neocortical neurons can change their firing properties from regular spiking to fast rhythmic bursting and fast spiking depending on afferent activity \citep{Steriade1998}. The maximum rate of cortical inhibitory interneurons is of the same order as that of pyramidal neurons (300--600 s$^{-1}$), and we set $Q_e^{\mathrm{max}} = Q_i^{\mathrm{max}} = 300$ s$^{-1}$ for simplicity (cf. Sec. \ref{sec:steady_states}). 

A maximum rate of $\sim$65 s$^{-1}$ was recorded for striatal neurons in awake, freely moving rats \citep{Kiyatkin1999}. High-frequency stimulation of the STN can evoke discharges up to about 200 s$^{-1}$ in GPi cells of rhesus monkeys \citep{Hashimoto2003}, while EP neurons were found to fire at rates up to 300 s$^{-1}$ in a rat slice preparation \citep{Nakanishi1990}. We take $Q_{p_1}^{\mathrm{max}} = 250$ s$^{-1}$ to be an adequate approximation. Of three types of neurons identified in the GP of guinea pigs, the most abundant type had a maximum firing rate close to 200 s$^{-1}$ \citep{Nambu1997}. On the other hand, \citet{Cooper2000} recorded the activity of three types of neuron in the rat GP with a weighted average maximum firing rate of $\sim$380 s$^{-1}$. We assume an intermediate value and let $Q_{p_2}^{\mathrm{max}} = 300$ s$^{-1}$. STN neurons in rats can fire at rates up to about 500 s$^{-1}$ \citep{Kita1983b,Nakanishi1987}. Finally, low-threshold Ca$^{2+}$ currents can cause thalamic neurons to fire high-frequency bursts at $\sim$300 s$^{-1}$ \citep{Destexhe2003}, while thalamic reticular neurons can fire bursts at up to 500 s$^{-1}$ \citep{Raeva1987}. 

The threshold values $\theta_a$ are the membrane depolarizations at which the populations fire at half their maximum rate [cf. Eq. (\ref{eq:1})]. Based on extensive exploration of physiologically realistic ranges, we choose values that give realistic steady-state firing rates for all neuronal populations in our model; these values are listed in Table~2. STN and pallidal neurons are taken to have low threshold potentials, while the threshold value is high for the relatively silent striatal neurons.
Note that for $Q \ll Q^{\mathrm{max}}$, a high maximum firing rate and a low threshold $\theta$ have closely similar effects on the (relatively low) steady-state firing rate, because
\begin{equation}\label{eq:7}
\frac{Q^{\mathrm{max}}}{1+ e^{-(V-\theta)/\sigma'}} \approx Q^{\mathrm{max}} e^{(V-\theta)/\sigma'},
\end{equation}
which means that adding $\delta\theta$ to $\theta$ is equivalent to replacing $Q^{\mathrm{max}}$ by $Q^{\mathrm{max}}e^{-\delta\theta/\sigma'}$. Hence, it is possible that the firing thresholds and maximum firing rates are both smaller or both larger than the values used here, leaving the dynamics largely unchanged. 

We also choose approximate connection strengths within physiological ranges, finding that the requirement of realistic firing rates restricts connection strengths to relatively narrow subranges. \citet{Robinson2004} derived values of $\nu_{es}$ and $\nu_{se}$ that satisfy both experimental constraints and give realistic firing rates in the purely corticothalamic model. Experiments in rodent neocortex in vitro \citep{Gil1999, Thomson1997} and in vivo \citep{Bruno2006} suggest that single thalamocortical and excitatory intracortical stimuli have a time-integrated response of about 10--20 $\mu$V s. However, $s_{es}$ should be adjusted for reduced probability of transmitter release at thalamocortical synapses upon repeated stimulation \citep{Gil1999}, and the less than additive effect of successive postsynaptic potentials \citep{Bruno2006}. If we assume the latter adjustment to yield an average unitary response of about 12 $\mu$V s, with a release probability of $40\%$ and $N_{es} = 85$ \citep{Bruno2006}, we obtain $\nu_{es}=0.4$ mV s. This value is in agreement with \citet{Robinson2004}, although that paper assumed a much smaller unitary response $s_{es}$ and a correspondingly larger number of synapses per cortical neuron. Studies of cat geniculocortical fibers suggest that the number of synapses of cortical origin per thalamic neuron far exceeds the number of thalamocortical synapses per cortical neuron \citep{Budd2004, Peters1993}. Corticothalamic fibers display paired-pulse facilitation, but often fail to evoke significant excitatory postsynaptic potentials (EPSPs) \citep{Golshani2001,Granseth2003}. Unitary corticothalamic EPSPs have a small amplitude compared to retinogeniculate EPSPs, with a time-integrated response around 12 $\mu$V s \citep{Turner1998}, similar to that at thalamocortical terminals. This combines with the large number of corticothalamic synapses but high failure rate to yield an approximate synaptic strength $\nu_{se} = 0.8$ mV s. We assume $\nu_{sr} = -0.4$ mV s, close to the value in \citet{Robinson2004}. Although there are fewer inhibitory than excitatory neurons in the cortex, the synaptic strength of the inhibitory projections is larger, such that $|\nu_{ei}|>|\nu_{ee}|$. 

Precise estimates of $\nu_{re}$ are difficult to obtain from the literature. Although it has been estimated that $\sim$60$\%$ of TRN terminals are of cortical origin \citep{Liu1999}, an estimate of the total number of synapses per TRN neuron is not readily available. In mouse brain slices, the unitary postsynaptic reponse of TRN neurons to cortical impulses is 2--3 times greater than that of thalamic relay neurons, and the latter have a high failure rate of transmitter release ($\sim$68$\%$), suggesting $s_{se} < s_{re}$. However, paired-pulse facilitation at relay neurons may partly counterbalance this effect \citep{Golshani2001}. More importantly, the relative influences of cortical inputs to relay and TRN neurons depend on the state of alertness \citep{Steriade1986}. During drowsiness or slow-wave sleep, TRN responds to cortical stimuli with bursts of spikes, implying a relatively large $\nu_{re}$, while TRN neurons are tonically active during attentive wakefulness and REM sleep, implying a lower value of $\nu_{re}$ \citep{Steriade2001b}. This is corroborated by modeling results, which indicate that the combined influence of the cortico-reticulo-thalamic and direct corticothalamic pathways is inhibitory in sleep states, but excitatory during waking \citep{Robinson2002}. We thus model a state of alert wakefulness in which the direct corticothalamic connection is more important than the cortico-reticulo-thalamic connection, setting $\nu_{re} = 0.15$ mV s. 

Although a high degree of convergence of corticostriatal inputs is suggested by the large number of synaptic events per unit time in striatal neurons \citep{Blackwell2003}, the strength of corticostriatal projections is constrained by the relatively low average firing rate of striatal neurons (cf. Sec. \ref{sec:rates}). Corticostriatal neurons form about 5000 -- 15 000 synapses per striatal cell, but each of these has a weak influence \citep{Wilson1995}. An immunohistochemical study found that the ipsilateral primary motor cortex innervated D1-expressing cells somewhat more strongly than D2-expressing cells \citep{Hersch1995}. We thus assume a strength per synapse of 0.1 $\mu$V s, with 10 000 synapses per striatal D1 cell, and 7000 synapses per D2 cell, which gives $\nu_{d_1e}=1.0$ mV s and $\nu_{d_2e}=0.7$ mV s. 

Steady states are very sensitive to $\nu_{sp_1}$ due to the high firing rate of neurons in the output nuclei. In order for the thalamus not to be overly suppressed by the output nuclei we assume a relatively small connection strength $\nu_{sp_1}=-0.03$ mV s. The thalamostriatal projection has been described as `massive' in the squirrel monkey \citep{Sadikot1992}, but this study did not quantify its strength or extent relative to other projections. Studies in rats suggest that it is much less important than the corticostriatal projection \citep{Groves1995}. \citet{Sidibe1996} found that intralaminar thalamic neurons preferentially innervate striatal neurons that output to GPi rather than GPe, so we let $\nu_{d_1s} = 0.1$ mV s and $\nu_{d_2s} = 0.05$ mV s. Evidence to suggest that $|\nu_{p_2d_2}| > |\nu_{p_1d_1}|$ comes from a study in which anterograde transport of a tracer injected into the arm area of the primary motor cortex labeled ten times as many GPe cells as GPi cells \citep{Strick1995}. A somewhat more symmetric distribution of striatal synapses over the two pallidal segments was suggested by \citet{Shink1995}, although the innervation of GPe still appeared stronger than that of GPi. Studies on the reciprocal connections between the GPe and STN suggest that GPe exerts a powerful inhibitory control over the STN \citep{Shink1996, Smith1990}, while the STN powerfully excites the GPe \citep{Cheruel1996, Nambu2000, Shink1996}. However, model results indicate that the GPe probably has a much weaker effect on the STN rate than vice versa, in view of stability criteria (cf. Paper II) and the comparatively high firing rate of the GPe (cf. Secs. \ref{sec:rates} and \ref{sec:firing_rate_results}). The STN appears to innervate both pallidal segments quite uniformly \citep{Parent1995, Smith1990b}, so we assume equal strengths for these projections. Fewer GPi synapses arise from the GPe than from the striatum, although the GPe may exert a strong effect by forming contacts with the perikarya and proximal dendrites of GPi cells \citep{Shink1995, Smith1994}. Still, we assume the direct pallidopallidal projection to be weaker than other projections, since a high GPe rate would otherwise be incompatible with an even higher GPi rate (cf. Secs. \ref{sec:rates} and \ref{sec:firing_rate_results}). Finally, given an input $\phi_n$, the connection strength for projections from brainstem to thalamus, $\nu_{sn}$, is constrained to a relatively narrow range since large values lead to unrealistically high cortical, thalamic, and striatal rates.

\subsection{Fixed points}\label{sec:steady_states}

Stable fixed points of the model equations, obtained assuming a constant input $\phi_n$, correspond to solutions at which the system settles down unless perturbed. Fixed-point equations are obtained by setting the derivatives in Eqs. (\ref{eq:2})--(\ref{eq:3}) to zero. Denoting fixed-point values by the superscript $(0)$, Eq. (\ref{eq:3}) yields
\begin{equation}\label{eq:4}
\phi_a^{(0)} = Q_a^{(0)},
\end{equation}
which, when substituted into Eq. (\ref{eq:2}) and using Eq. (\ref{eq:1}), gives
%
%
\begin{equation}\label{eq:5}
S_a^{-1}\left(\phi_a^{(0)}\right) = \sum_b \nu_{ab}\phi_b^{(0)}.
\end{equation}
Implementing Eq. (\ref{eq:5}) for all nine neuronal populations gives a set of transcendental equations for the fixed-point firing rates $\phi_a^{(0)}$ that can be solved numerically. These equations are simplified by imposing the \emph{random connectivity approximation}, in which the number of synapses within the cortex is proportional to the product of the numbers of sending and receiving neurons \citep{Braitenberg1998, Robinson2001, Wright1995}. In the connection strengths $\nu_{ab} = N_{ab}s_{ab}$, the symbol $N_{ab}$ denotes the number of synapses from neurons of type $b$ \emph{per type $a$ neuron}, which in the random connectivity approximation thus depends only on the afferent population $b$. If we further assume the unitary synaptic strengths $s_{ab}$ to be independent of the receiving population, we obtain
\begin{equation}\label{eq:6}
\nu_{ee} = \nu_{ie}, \ \ \ \nu_{ei}=\nu_{ii}, \ \ \ \nu_{es}=\nu_{is}.
\end{equation}
For $Q_e^{\mathrm{max}} = Q_i^{\mathrm{max}}$ and $\theta_e = \theta_i$ (cf. Sec. \ref{sec:parameters}) this implies, in particular, that the fixed-point values of the cortical excitatory and inhibitory firing rate fields are equal, since identical equations are obtained for $\phi_e^{(0)}$ and $\phi_i^{(0)}$.

\begin{figure}[htbp]
\centering
\includegraphics[width=430pt]{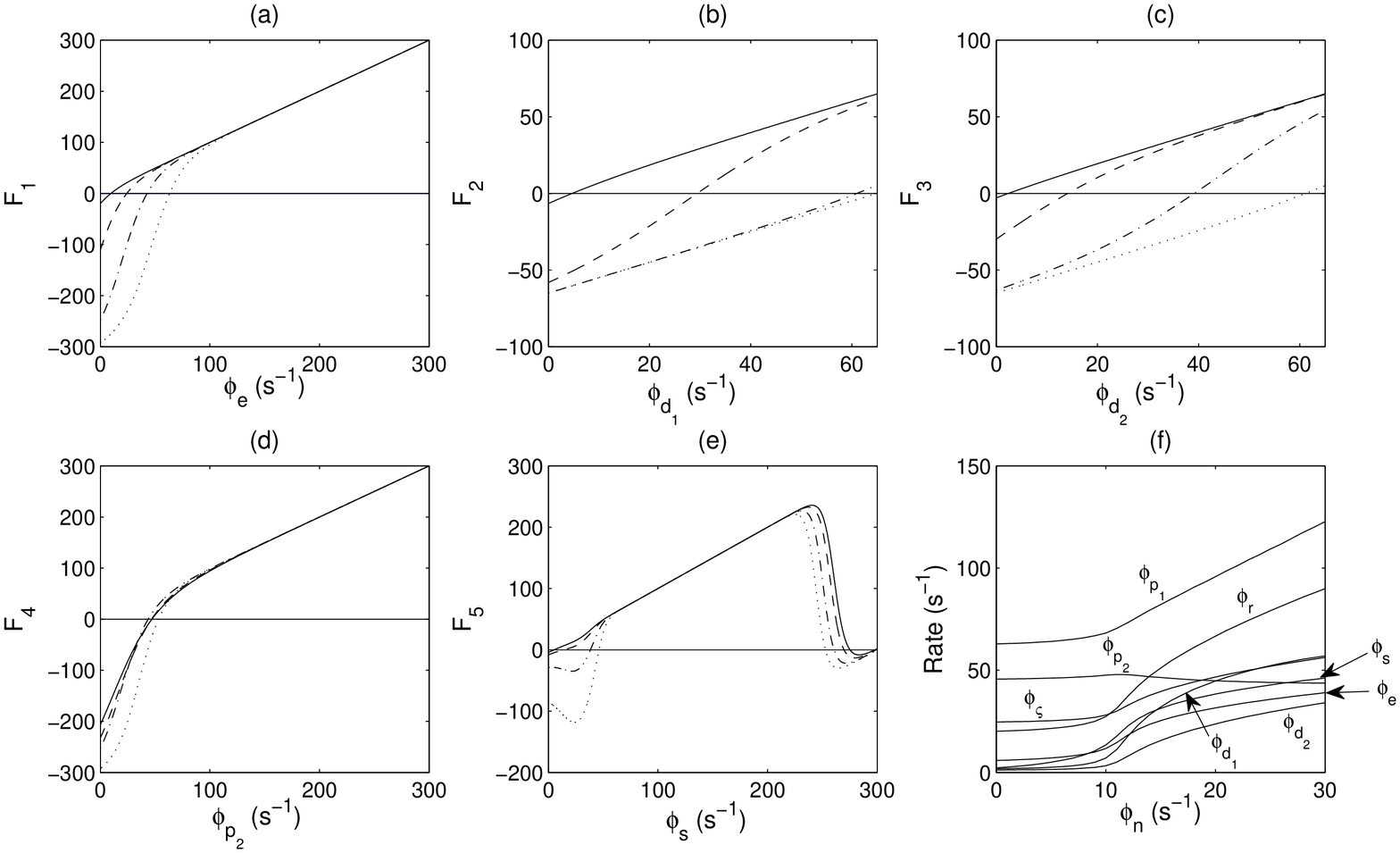}
\caption{Equilibrium firing rates of the BGTCS. (a)--(d) Equilibrium values of $\phi_e$, $\phi_{d_1}$, $\phi_{d_2}$, and $\phi_{p_2}$ (zeros of $F_1-F_4$) for different thalamic rates $\phi_s$. Solid, $\phi_s = 10$ s$^{-1}$; dashed, $\phi_s = 30$ s$^{-1}$; dash-dotted, $\phi_s = 50$ s$^{-1}$; dotted, $\phi_s = 70$ s$^{-1}$. (e) Equilibrium values of $\phi_s$ (zeros of $F_5$) for different external inputs $\phi_{n}$. Solid, $\phi_n = 5$ s$^{-1}$; dashed, $\phi_n = 10$ s$^{-1}$; dash-dotted, $\phi_n = 20$ s$^{-1}$; dotted, $\phi_n = 30$ s$^{-1}$. (f) Dependence of low-firing-rate fixed points on $\phi_n$.}
\label{fig:steady_state_plots}
\end{figure}

In practice, fixed points can be determined by considering the simultaneous zeros of the five functions
\begin{align}
F_1(\phi_e) = \phi_e - S_e[(\nu_{ee}+\nu_{ei})\phi_e + \nu_{es}\phi_s], \\
F_2(\phi_{d_1}) = \phi_{d_1} - S_{d_1}(\nu_{d_1e}\phi_e + \nu_{d_1d_1}\phi_{d_1} + \nu_{d_1s}\phi_s), \\
F_3(\phi_{d_2}) = \phi_{d_2} - S_{d_2}(\nu_{d_2e}\phi_e + \nu_{d_2d_2}\phi_{d_2}+ \nu_{d_2s}\phi_s), \\
F_4(\phi_{p2}) = \phi_{p_2} - S_{p_2}[\nu_{p_2d_2}\phi_{d_2} + \nu_{p_2p_2}\phi_{p_2} + \nu_{p_2\varsigma}S_{\varsigma}(\nu_{\varsigma e}\phi_e + \nu_{\varsigma p_2}\phi_{p_2})],\\
F_5(\phi_s) = \phi_s - S_s(\nu_{se}\phi_e + \nu_{sp_1}\phi_{p_1} + \nu_{sr}\phi_r + \nu_{sn}\phi_{n}).\label{eq:F5}
\end{align}
First an initial estimate of the thalamic firing rate $\phi_s$ is made. For the given choice of $\phi_s$, the cortical excitatory rate $\phi_e$ is uniquely determined by the zero of $F_1$, as shown in Fig. \ref{fig:steady_state_plots}(a) using the parameter values in Table~2. From this the striatal rates $\phi_{d_1}$ and $\phi_{d_2}$ can be determined using the functions $F_2$ and $F_3$. These are negative at $\phi_{d_1}, \phi_{d_2} = 0$ and positive at the maximum striatal firing rates, crossing zero exactly once since $\nu_{d_1d_1} < 0$ and $\nu_{d_2d_2} < 0$ ensure that their derivatives
\begin{eqnarray}
\frac{dF_2(\phi_{d_1})}{d\phi_{d_1}} &=& 1 - \nu_{d_1d_1} Q_{d_1}^{\mathrm{max}} \frac{ \exp[-(\nu_{d_1e}\phi_e +\nu_{d_1d_1}\phi_{d_1} + \nu_{d_1s}\phi_s - \theta_{d_1})/\sigma'] }{\sigma'\{1+\exp[-(\nu_{d_1e}\phi_e +\nu_{d_1d_1}\phi_{d_1} + \nu_{d_1s}\phi_s - \theta_{d_1})/\sigma' ]\}^2},\\
\frac{dF_3(\phi_{d_2})}{d\phi_{d_2}} &=& 1 - \nu_{d_2d_2} Q_{d_2}^{\mathrm{max}} \frac{\exp[-(\nu_{d_2e}\phi_e +\nu_{d_2d_2}\phi_{d_2} + \nu_{d_2s}\phi_s - \theta_{d_2})/\sigma'] }{\sigma'\{1+\exp[-(\nu_{d_2e}\phi_e +\nu_{d_2d_2}\phi_{d_2} + \nu_{d_2s}\phi_s - \theta_{d_2})/\sigma' ]\}^2},
\end{eqnarray}
are always positive. The functions $F_2$ and $F_3$ are shown in Figs. \ref{fig:steady_state_plots}(b) and \ref{fig:steady_state_plots}(c). The striatal rates increase with both the thalamic rate $\phi_s$ and the cortical rate $\phi_e$. The unique zero of the function $F_4$ determines the GPe rate $\phi_{p_2}$. The value of $\phi_{p_2}$ first decreases, then increases slightly with both $\phi_s$ and $\phi_e$ [Fig. \ref{fig:steady_state_plots}(d)]. Knowing $\phi_{p_2}$ in turn allows determination of the STN and GPi/SNr rates $\phi_{\varsigma}$ and $\phi_{p_1}$, and finally, the self-consistency relation $F_5 = 0$ must be satisfied for $\phi_s$ to represent a fixed point. The value of $F_5$ is negative at $\phi_s = 0$, and positive at $\phi_s = Q_s^{\mathrm{max}}$, since the last term in (\ref{eq:F5}) is always smaller than $Q_s^{\mathrm{max}}$. By continuity of $F_5$, there is always at least one fixed point, and in general an odd number of fixed points. The function $F_5$ plotted in Fig. \ref{fig:steady_state_plots}(e) shows that there are three fixed points for the parameters in Table~2. 
Since the low-firing-rate fixed point for $\phi_s$ is stable and yields the most realistic firing rates for all populations, we will take this fixed point to represent the physiological situation. It is important to note that we obtain realistic steady states with parameter values that are consistent with experimental findings (cf. Sec. \ref{sec:parameters}); this is a fundamental test of the physiological realism of the model. Figs. \ref{fig:steady_state_plots}(e) and \ref{fig:steady_state_plots}(f) show that all firing rates increase with the brainstem input $\phi_n$.

\subsection{Modeling dopamine depletion}\label{sec:depletion}

It is widely recognized that dopamine modulates corticostriatal transmission both presynaptically and postsynaptically \citep{Calabresi2000}. However, observations on these effects are complicated and sometimes paradoxical. For instance, dopamine may facilitate glutamate-induced activity at low concentrations, but be inhibitory at higher concentrations \citep{Hu1997}. Joint activation of D1 and D2 receptors may have a supra-additive effect on striatal neurons, and upregulation of both receptor types after dopamine depletion may influence corticostriatal transmission \citep{Hu1990}. 

Nevertheless, some findings are relatively consistent across studies. Dopamine appears to have a predominantly inhibitory effect via D1 receptors when striatal neurons are in a hyperpolarized state \citep{Nicola1998}, but a facilitatory effect when they are already in an activated state \citep{Hernandez-Lopez1997, Nicola2000}. Such facilitation via D1 activation may primarily affect NMDA receptor-mediated transmission \citep{Cepeda1998,Levine1996}. In line with these observations, some researchers have suggested that dopamine increases the signal-to-noise ratio (SNR) in the striatum, making medium spiny neurons sensitive only to strong inputs \citep{Nicola2004,O'Donnell2003}. This is supported by the finding that dopamine reduces the basal level of striatal activity, but enhances the phasic response to glutamate \citep{Kiyatkin1996}. To model the reduction in SNR following loss of dopamine, \citet{Leblois2006} suggested decreasing the firing thresholds of striatal neurons as well as corticostriatal connection strengths to both D1- and D2-expressing neurons. They assumed both thresholds and connection strengths to depend nearly linearly on the level of dopamine. We use
\begin{eqnarray}
\theta_{d_1}^{\mathrm{new}} = \theta_{d_1}-h\chi;\hspace{1cm} \theta_{d_2}^{\mathrm{new}} = \theta_{d_2}-h\chi, \label{eq:SNR1}\\
\nu_{d_1e}^{\mathrm{new}} = \nu_{d_1e} - \chi;\hspace{1cm} \nu_{d_2e}^{\mathrm{new}} = \nu_{d_2e} - \chi,\label{eq:SNR2}
\end{eqnarray} 
where we consider $h =$ 5, 10, and 15 s$^{-1}$, and $\chi$ between 0 and 0.6 mV s. Note that the effective thresholds of D1 and D2 neurons are approximated as equal for simplicity, although their intrinsic properties may differ in practice \citep{Moyer2007}.

In contrast to the above possibility, dopamine loss may also reduce transmission via D1-expressing cells and enhance transmission via D2-expressing cells. In support of this possibility, a number of studies have shown that depression of excitation via glutamatergic, particularly AMPA, receptors follows activation of D2 receptors \citep{Hsu1995,Levine1996,Toan1985,Umemiya1997}. Furthermore, \citet{Mallet2006} found an increase in both the spontaneous activity and responsiveness of striatopallidal neurons after dopamine depletion, whereas striatonigral neurons became less active. Mallet and coworkers suggested that this imbalance may be exacerbated by feedforward inhibition by fast-spiking GABAergic interneurons, which narrow the time window for integration of cortical inputs in striatonigral neurons, and widen the time window in striatopallidal neurons, leading to inappropriate summation of signals in the indirect pathway. This possibility can be modeled by simultaneously increasing the strength of corticostriatal projections giving rise to the indirect pathway, and decreasing that of projections giving rise to the direct pathway. In the absence of detailed information concerning the relative sizes of the changes in these connection strengths, we decrease $\nu_{d_1e}$ and increase $\nu_{d_2e}$ by the same factor $\xi$. We also consider the effects of modulating $\nu_{d_1e}$ and $\nu_{d_2e}$ independently. Of course, the degree of differential modulation of the direct and indirect pathways by dopamine depletion is limited by the extent of segregation of D1 and D2 class receptors, and the separation of striatal projections to GPi/SNr and GPe. Nevertheless, differential modulation may occur with only a partial distinction between the direct and indirect pathways. 

Reduced release of dynorphin and enhanced release of enkephalin in the striatum are thought to represent compensatory mechanisms that oppose the effects of chronic dopamine depletion \citep{Augood1989,Betarbet2004,Engber1992}. For instance, dynorphin appears to inhibit the release of dopamine in the striatum, and oppose the effects of D1-receptor stimulation on striatonigral neurons \citep{Steiner1998}. Therefore, the loss of dynorphin will enhance the sensitivity of D1-expressing neurons to the remaining dopamine. Dynorphin and enkephalin may also affect striatal input to the GPe as well as intrapallidal inhibition \citep{Ogura2000,Stanford1999}. Enkephalin appears to act presynaptically via opioid receptors to inhibit the release of GABA at both striatopallidal and intrapallidal terminals \citep{Maneuf1994,Stanford1999}, decreasing the corresponding connection strengths. These effects may be partly opposed by loss of dynorphin from axon collaterals of neurons projecting to both pallidal segments \citep{Ogura2000}. Furthermore, the loss of direct dopaminergic afferentation promotes GABA release at striato-GPe terminals through the action of presynaptic D2 receptors \citep{Floran1997, Querejeta2001}. Assuming that striatopallidal transmission is more affected by dopamine loss than by enkephalin, we model PD with an increase in striato-GPe inhibition. The prevalence of enkephalin over dynorphin in the GPe suggests that the strength of intrapallidal inhibition is reduced. This agrees with the suggestion of \citet{Terman2002} that dopaminergic denervation weakens the lateral connections in the GPe, although they assumed enkephalin and dynorphin to exert concerted rather than opposing effects. Reduced dynorphin levels may also depolarize GPe neurons by blocking membrane potassium conductance \citep{Ogura2000}, which we model by lowering the corresponding threshold potential.

Several lines of evidence indicate that STN overactivity is not only caused by reduced GPe firing. For instance, GPe lesions only cause a slight increase in STN rate \citep{Hassani1996}, and blockade of glutamatergic transmission suppresses STN overactivity in rats with haloperidol-induced akinesia \citep{Miwa1998}. Increased enkephalin levels may inhibit GABAergic synaptic transmission via $\mu$-opioid receptors, which are expressed in high concentration in the human STN \citep{Peckys1999,Raynor1995}, although enkephalin also suppresses excitatory transmission \citep{Shen2002}. Studies in rats have suggested that enhanced excitation by the thalamic parafascicular nucleus \citep{Orieux2000} and/or the pedunculopontine nucleus (PPN) \citep{Breit2001, Breit2006, Orieux2000} may contribute to STN hyperactivity in parkinsonism. Rats treated with 6-OHDA have elevated levels of extracellular potassium in the STN, possibly due to changes in conductivity and delayed clearance, which increases the activity of this nucleus \citep{Strauss2008}. Furthermore, the extracellular concentration of glutamate is increased \citep{Fujikawa1996}, and that of GABA decreased \citep{Engblom2003}, by higher extracellular levels of potassium. We model the hyperexcitability of STN neurons in PD by reducing their average threshold potential.

The frontal lobe, including the prefrontal cortex, SMA, and M1, receives a significant dopaminergic innervation from the SNc, VTA, and retrorubral area \citep{Gaspar1992, Williams1993}. The predominant influence of dopamine on prefrontal pyramidal cell firing is inhibitory, due to enhanced synaptic inputs from GABAergic interneurons \citep{Gulledge2001, Sesack1989,Zhou1999}. \citet{Gulledge2001} suggested that this is because of enhanced GABA release by interneurons independent of their spike rate, implying increased synaptic strengths $|\nu_{ei}|$ and $|\nu_{ii}|$. 
Dopamine also renders interneurons more susceptible to excitatory inputs from pyramidal cells, apparently without affecting their firing threshold or the amplitude of EPSPs \citep{Gao2003}. This mechanism has no precise equivalent in our model, but can be approximated by an increased synaptic strength $\nu_{ie}$. Furthermore, changes in intrinsic currents increase the excitability of pyramidal neurons of the rat prefrontal cortex after D1 receptor activation in vitro \citep{Thurley2008}, which may positively affect $\nu_{ee}$. This suggests that we can model PD in part by decreasing the synaptic strengths $\nu_{ee}$, $\nu_{ie}$, and especially $|\nu_{ei}|$ and $|\nu_{ii}|$, in accord with the reduced intracortical inhibition observed upon transcranial magnetic stimulation in PD patients off medication \citep{Ridding1995}.

We thus model the effects of nigrostriatal degeneration in the following five ways:\\
(I) mimicking a decrease in striatal SNR by reducing both firing thresholds and corticostriatal connection strengths according to Eqs. (\ref{eq:SNR1}) and (\ref{eq:SNR2});\\
(II) increasing $\nu_{d_2e}$ and decreasing $\nu_{d_1e}$ either individually or by the same factor $\xi$, in agreement with the direct/indirect pathway model;\\
(III) reducing lateral inhibition in the GPe, which mimicks enhanced levels of enkephalin;\\
(IV) attenuating cortical interactions to capture loss of intrinsic cortical dopamine; \\
(V) a combination of a stronger indirect and weaker direct pathway ($\nu_{d_1e}=0.5$ mV s, $\nu_{d_2e}=1.4$ mV s), reduced intrapallidal inhibition ($\nu_{p_2p_2}=-0.07$ mV s), weaker intracortical coupling ($\nu_{ee}=\nu_{ie}=1.4$ mV s, $\nu_{ei}=\nu_{ii}=-1.6$ mV s), lower STN and GPe firing thresholds ($\theta_{p_2}=8$ mV, $\theta_{\varsigma}=9$ mV), and a stronger striato-GPe projection ($\nu_{p_2d_2}=-0.5$ mV s). We will call the state corresponding to these parameters the `full parkinsonian state'.  

\section{Results}\label{sec:firing_rate_results}

We now describe the results of applying the model equations to the system of connections depicted in Fig. \ref{fig:diagram}. The steady-state firing rates corresponding to the parameter values in Table~2, with a stimulus level of 10 s$^{-1}$, are listed in column $a$ of Table~3. All rates are in physiologically realistic ranges for healthy individuals if the average striatal rate is taken to be $(\phi_{d_1}+\phi_{d_2})/2$ (cf. Table~1). Changes in firing rates are derived when modeling dopamine loss in the five ways listed in Sec. \ref{sec:depletion}. Model predictions are compared with the empirical findings summarized in Sec. \ref{sec:rates}. \\

\begin{table}[htp]
\centering
\begin{minipage}{\textwidth} 
\centering
\begin{tabular}{p{1.9cm}p{0.45cm}p{0.45cm}p{0.45cm}p{0.45cm}p{0.45cm}p{0.45cm}p{0.45cm}p{0.45cm}p{0.45cm}p{0.45cm}p{0.45cm}p{0.45cm}p{0.45cm}p{0.45cm}p{0.45cm}}
\hline 
 & $a$ & $b$ & $c$ & $d$ & $e$ & $f$ & $g$ & $h$ & $i$ & $j$ & $k$ & $l$ & $m$ & $n$ & $o$\\
\hline
Cortex & 12 & 12 & 10 & 16 & 12 & 22 & 14 & 12 & 13 & 12 & 13  & 11 & 10 & 14 & 10 \\
D1 & 7.4 & 6.0 & 1.9 & 14 & 2.4 & 24 & 2.8 & 2.2 & 15 & 2.6 & 13 & 6.7 & 5.0 & 11 & 5.1  \\
D2 & 3.5 & 2.7 & 9.3 & 6.4 & 12 & 11 & 16 & 12 & 5.4 & 24 & 3.9 & 5.8 & 2.5 & 4.9 & 2.6  \\
GPi/SNr & 69 & 69 & 83 & 49 & 70 & 78 & 100 & 110  & 63 & 110 & 64 & 72 & 87 & 56 & 85  \\
GPe & 48 & 48 & 40 & 65 & 51 & 48 & 36 & 47 & 46 & 28 & 48 & 45 & 47 & 58 & 55 \\
STN & 28 & 28 & 29 & 27 & 27 & 36 & 33 & 36 & 29 & 34 & 29  & 29 & 27 & 27 & 32\\
Relay nuclei & 14 & 14 & 11 & 20 & 13 & 22 & 13 & 10 & 15 & 10 & 15 & 13 & 10 & 17 & 11 \\
TRN & 28 & 28 & 25 & 34 & 27 & 42 & 29 & 27 & 29 & 27 & 29  & 27 & 25 & 31 & 25\\
\hline
\end{tabular}
\end{minipage}
\caption{Firing rates of the components of the BGTCS (in s$^{-1}$) corresponding to the following cases: (a) healthy state, as represented by the parameters in Table 2; (b) reduced-SNR state, obtained from $a$ by setting $h=10$ s$^{-1}$, $\chi=0.6$ mV s, which leads to $\theta_{d_1}=\theta_{d_2}=13$ mV, $\nu_{d_1e}=0.4$ mV s, $\nu_{d_2e}=0.1$ mV s; (c) state $a$ with a stronger indirect and weaker direct pathway, $\nu_{d_1e} = 0.5$ mV s and $\nu_{d_2e} = 1.4$ mV s; (d) state $a$ with weaker intrapallidal inhibition $\nu_{p_2p_2}=-0.03$ mV s; (e) state $c$ with $\nu_{p_2p_2}=-0.03$ mV s; (f) state $a$ with weaker cortical interactions, $\nu_{ee}=\nu_{ie}=1.4$ mV s, $\nu_{ei}=\nu_{ii}=-1.6$ mV s; (g) state $c$ with $\nu_{ee}=\nu_{ie}=1.4$ mV s, $\nu_{ei}=\nu_{ii}=-1.6$ mV s; (h) full parkinsonian state' (cf. Sec. 3.4); (i) state $a$ with $\nu_{d_1d_1}=\nu_{d_2d_2}=0$ mV s; (j) state $g$ with $\nu_{d_1d_1}=\nu_{d_2d_2}=0$ mV s; (k) state $a$ with $\nu_{d_1s}=0.3$ mV s; (l) state $a$ with $\nu_{d_2s}=0.3$ mV s; (m) state $a$ with $\nu_{p_1p_2}=0$ mV s; (n) state $a$ with $\nu_{p_2\varsigma}=0.4$ mV s; (o) state $a$ with $\nu_{\varsigma e}=0.2$ mV s. Input to the thalamic relay nuclei is 10 s$^{-1}$ and all rates are given to two significant figures.}
\end{table}

\noindent \emph{Effects of parameter changes mimicking a reduced signal-to-noise ratio}\\

\begin{figure}[htp]
\centering
   \includegraphics[width=250pt, height=210pt]{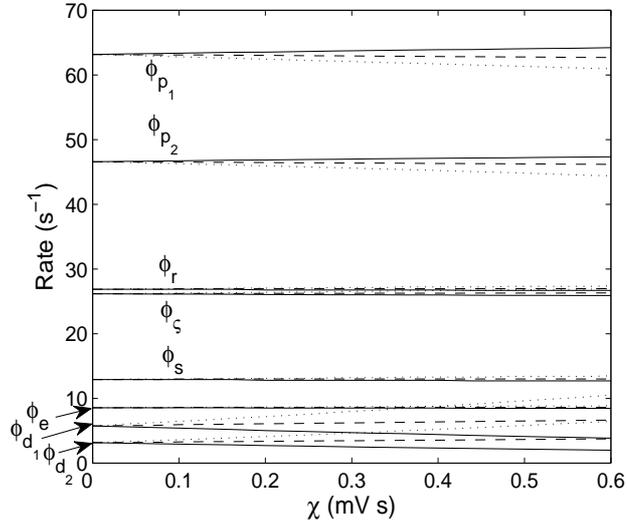}
\caption{Changes in firing rates due to reduction in striatal SNR. The parameter $h$ in Eq. (\ref{eq:SNR1}) is taken to be either 5 s$^{-1}$ (solid), 10 s$^{-1}$ (dashed), or 15 s$^{-1}$ (dotted). Groups of lines start together at $\chi=0$ mV s.}
\label{fig:SNR}
\end{figure}

Irrespective of the relative sizes of the changes in striatal firing thresholds and corticostriatal connection strengths, a reduction in SNR simulated by increasing $\chi$ [cf. Eqs. (\ref{eq:SNR1}) and (\ref{eq:SNR2})] has little impact on average firing rates (cf. Fig. \ref{fig:SNR}). Representative rates are listed for $h=10$ s$^{-1}$ and $\chi=0.6$ mV s in Column $b$ of Table~3. In the model of \citet{Leblois2006}, which lacks an indirect pathway but in which dopamine loss is also modeled by approximating a reduced SNR, resting activity in the GPi also changes little with a lower dopamine level, whereas cortical activity shows a small increase. The model of \citet{Leblois2006} is not directly comparable to ours, since it contains a different set of connections, forming separate streams in the direct path but with diffuse projections from STN to GPi in the hyperdirect pathway. Furthermore, it considers individual neurons with linear response functions above threshold, a different form of the dendritic/synaptic filter function, and identical firing thresholds for different neurons in each given population except the striatal one, whereas the sigmoid function (\ref{eq:1}) results from a distribution of firing thresholds. However, their (deliberately) small change in average activity with a reduced SNR corresponds well with our result. \\

\noindent \emph{Effects of stronger indirect and weaker direct pathway}\\

\begin{figure}[htp]
\centering
\includegraphics[width=430pt]{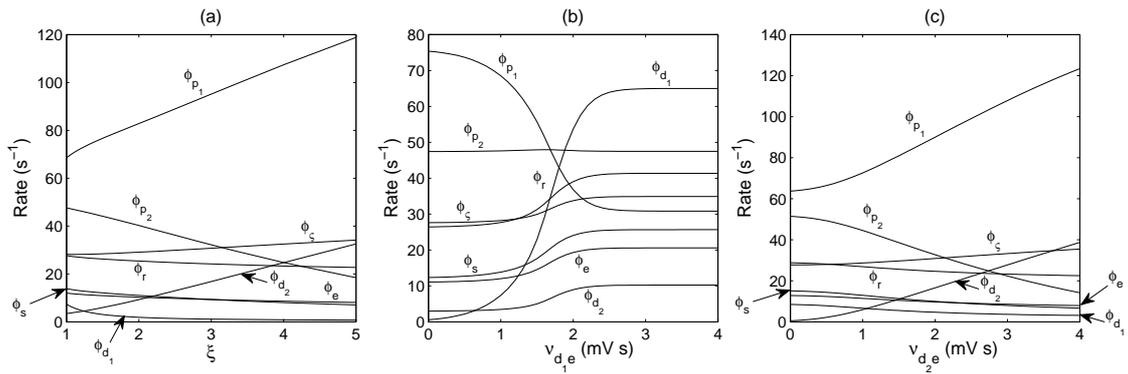}
\caption{Changes in firing rates when modeling dopamine loss by an increase in $\nu_{d_2e}$ and decrease in $\nu_{d_1e}$. (a) Rates when multiplying $\nu_{d_2e}$ and dividing $\nu_{d_1e}$ by the same factor $\xi$. (b) Rates vs. $\nu_{d_1e}$. (c) Rates vs. $\nu_{d_2e}$.}
\label{fig:corticostriatal}
\end{figure}

A simultaneous increase in $\nu_{d_2e}$ and decrease in $\nu_{d_1e}$ by an identical factor $\xi$ leads to changes in firing rates that agree well with the average of a cross-section of empirical findings [cf. Fig. \ref{fig:corticostriatal}(a) and Table~1]: GPe and thalamic rates are reduced, while STN and GPi/SNr rates are elevated. Because the increase in D2 rate is greater than the reduction in D1 rate, the overall striatal rate is slightly increased. Figure \ref{fig:corticostriatal}(b) and \ref{fig:corticostriatal}(c) show that these changes are effected mainly through the indirect pathway, since a weaker $\nu_{d_1e}$ decreases $\phi_{\varsigma}$, $\phi_{d_1}$, and $\phi_{d_2}$, and barely affects $\phi_{p_2}$. Doubling the strength of the indirect pathway while reducing that of the direct pathway by the same factor (\mbox{$\nu_{d_1e} = 0.5$ mV s} and \mbox{$\nu_{d_2e} = 1.4$ mV s}) leads to the rates in Column $c$ of Table~3. 

The sizes of these changes are fairly realistic, with a relatively large change in the rate of the output nuclei and smaller changes in cortical, thalamic, and average (over the D1 and D2 populations) striatal rates (cf. Sec. \ref{sec:rates}), but the increase in STN activity is smaller than would be expected on the basis of empirical evidence, while the decrease in GPe activity is quite large. Although our model predicts that GPe lesion increases the rate of the remaining intact GPe neurons (cf. Paper II), lesions are nevertheless expected to decrease the total GPe output. Therefore, our finding that a relatively large decrease in GPe output only slightly enhances STN firing accords with the results of GPe lesion reported by \citet{Hassani1996}, and confirms the influence of other excitatory mechanisms on the STN rate mentioned in Sec. \ref{sec:depletion}.\\

\newpage
\noindent \emph{Effects of weakened intrapallidal inhibition}\\

\begin{figure}[htp]
\centering
\includegraphics[width=430pt]{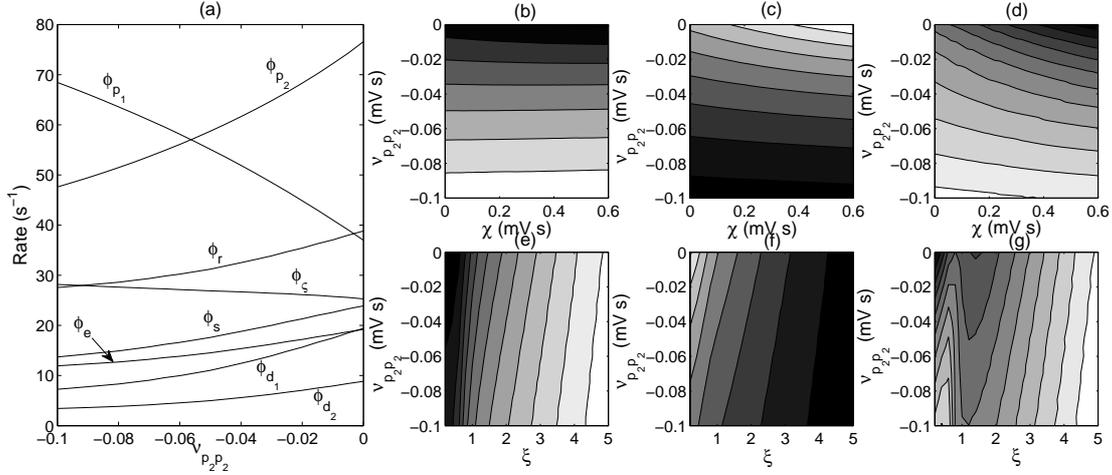}
\caption{Sensitivity of steady-state firing rates to lateral inhibition in the GPe. (a) Changes in pallidal and STN rates with smaller $|\nu_{p_2p_2}|$ are opposite to those in parkinsonism. (b)--(d) From left to right: contour plots of GPi/SNr, GPe, and STN rates as functions of $\nu_{p_2p_2}$ and striatal SNR, where $h=10$ s$^{-1}$ [cf. Eqs. (\ref{eq:SNR1}) and (\ref{eq:SNR2})]. (e)--(g) From left to right: contour plots of GPi/SNr, GPe, and STN rates as functions of $\nu_{p_2p_2}$ and the relative strengths of the direct and indirect pathways. Lighter shades correspond to higher rates.}
\label{fig:GPe_sensitivity}
\end{figure}

Figure \ref{fig:GPe_sensitivity}(a) shows the results of weakened lateral inhibition in the GPe. The effects on the pallidal and STN rates are opposite to those expected in PD, in line with the presumed compensatory effects of increased enkephalin levels with chronic dopamine depletion. Figures \ref{fig:GPe_sensitivity}(b)--(d) illustrate how a combination of weakened intrapallidal inhibition and a reduced SNR [$h=10$ s$^{-1}$ in Eq. (\ref{eq:SNR1})] affects the GPi/SNr, GPe, and STN rates, respectively. The corresponding results for a weakened direct and strengthened indirect pathway are shown in Figs. \ref{fig:GPe_sensitivity}(e)--(g). The reduction in $|\nu_{p_2p_2}|$ causes parkinsonian rates to be reached more slowly with an increase in $\xi$. Column $d$ of Table~3 contains the steady-state firing rates for $\nu_{p_2p_2}=-0.03$ mV s, where $\chi=0$ mV s and $\xi=1$. The rates for $\nu_{d_1e}=0.5$ mV s, $\nu_{d_2e}=1.4$ mV s, and $\nu_{p_2p_2}=-0.03$ mV s are given in Column $e$.\\

\noindent \emph{Effects of weakened cortical interactions}\\

\begin{figure}[htp]
\centering
\includegraphics[width=400pt, height=170pt]{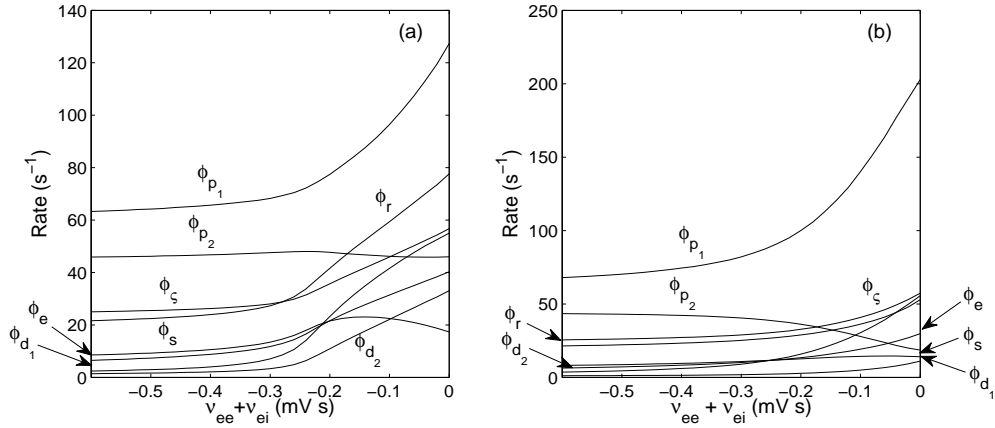}
\caption{Sensitivity of steady-state firing rates to intracortical connection strengths. Rates only depend on the sum $\nu_{ee} + \nu_{ei}$ due to the random connectivity approximation (cf. Sec. \ref{sec:steady_states}). (a) Variations with respect to the healthy state. (b) Variations with respect to the state with $\nu_{d_1e}=0.5$ mV s, $\nu_{d_2e}=1.4$ mV s.}
\label{fig:cortical_sensitivity}
\end{figure}

Loss of intrinsic cortical dopamine due to degeneration of the mesocortical pathway is expected to reduce intracortical coupling, and especially the strength of inhibition. Due to the random connectivity approximation (cf. Sec. \ref{sec:steady_states}), steady-state firing rates depend only on the sum $\nu_{ee} + \nu_{ei}$, rather than on the individual connection strengths. The sensitivity of the firing rates of the various components to this sum is depicted in Fig. \ref{fig:cortical_sensitivity}. It is seen that loss of mesocortical dopamine helps to normalize the cortical rate after nigrostriatal damage, and further increases the STN rate. The effect on the GPe rate depends on the relative strengths of the direct and indirect pathways: it remains almost constant with mesocortical dopamine loss if the SNc is intact, but decreases with mesocortical dopamine loss after SNc lesion (modeled with $\nu_{d_1e}=0.5$ mV s and $\nu_{d_2e}=1.4$ mV s). The rates for $\nu_{ee} = \nu_{ie} = 1.4$ mV s and $\nu_{ei}=\nu_{ii}=-1.6$ mV s are listed in Column $f$ of Table~3, with the corresponding values where also $\nu_{d_1e}=0.5$ mV s and $\nu_{d_2e}=1.4$ mV s in Column $g$. \\ 

\noindent \emph{Effects of GPe and STN firing thresholds and the striato-GPe projection}\\

\begin{figure}[htp]
\centering
\includegraphics[width=400pt, height=170pt]{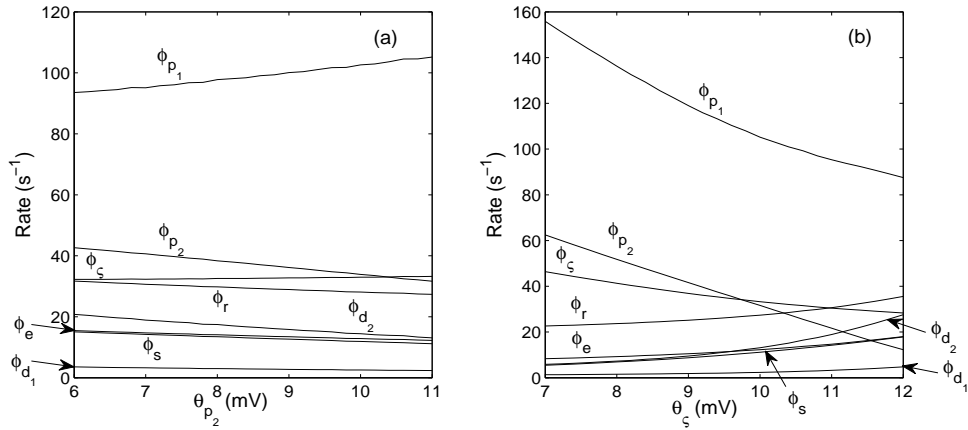}
\caption{Sensitivity of steady-state firing rates to STN and GPe firing thresholds, where $\nu_{d_1e}=0.5$ mV s, $\nu_{d_2e}=1.4$ mV s, $\nu_{ee}=\nu_{ie}=1.4$ mV s, and $\nu_{ei}=\nu_{ii}=-1.6$ mV s to mimick nigrostriatal and mesocortical dopamine depletion. (a) Dependence on $\theta_{p_2}$. (b) Dependence on $\theta_{\varsigma}$.}
\label{fig:threshold_sensitivity}
\end{figure}

Besides the above changes, nigrostriatal degeneration may lead to reduced STN and GPe firing thresholds and an increase in $|\nu_{p_2d_2}|$, as discussed in Sec. \ref{sec:depletion}. Changes in firing rates with $\theta_{p_2}$ and $\theta_{\varsigma}$ are shown in Fig. \ref{fig:threshold_sensitivity}, where nigrostriatal and mesocortical dopamine loss are taken into account via $\nu_{d_1e}=0.5$ mV s, $\nu_{d_2e}=1.4$ mV s, $\nu_{ee}=\nu_{ie}=1.4$ mV s, and $\nu_{ei}=\nu_{ii}=-1.6$ mV s. A reduction in $\theta_{p_2}$ normalizes all rates that were altered by SNc lesion, except that of striatal D2 cells. A smaller $\theta_{\varsigma}$ has the opposite effect apart from increasing the GPe rate. Together, lower STN and GPe firing thresholds limit the decrease in GPe rate, counterbalance the increase in corticothalamic and striatal rates caused by mesocortical dopamine loss, and help account for a relatively large increase in STN rate. 

\begin{figure}[htp]
\centering
\includegraphics[width=430pt]{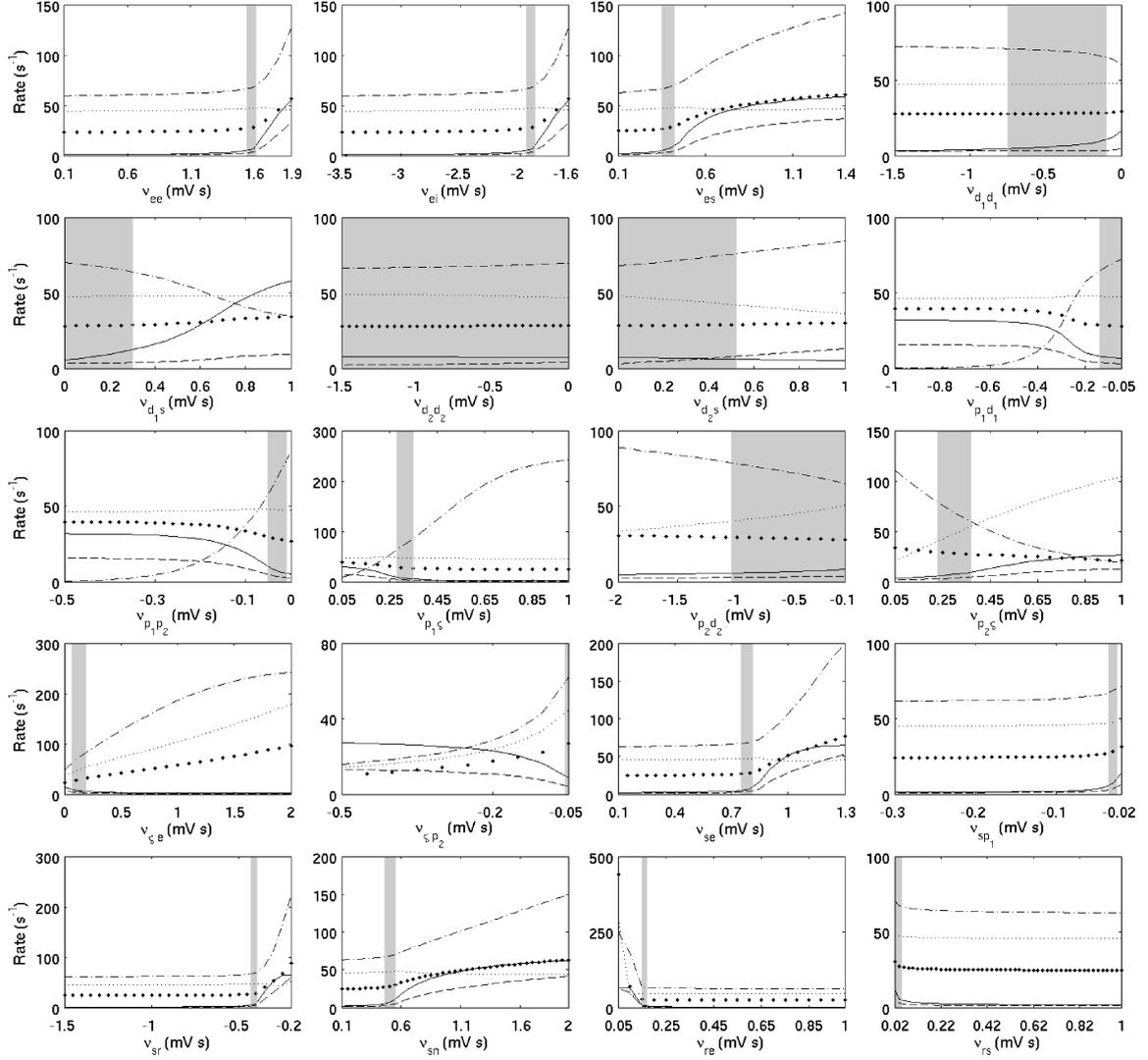}
\caption{Sensitivity of steady-state firing rates to connection strengths $\nu_{ab}$. Shaded regions indicate parameter ranges that yield realistic firing rates in the healthy state for all neuronal populations. Solid, $\phi_{d_1}$; dashed, $\phi_{d_2}$; dash-dotted, $\phi_{p_1}$, dotted, $\phi_{p_2}$; large dots, $\phi_{\varsigma}$.}
\label{fig:sensitivity}
\end{figure}

Figure \ref{fig:sensitivity} shows the sensitivity of the steady-state firing rates to the remaining connection strengths, with all other parameters held constant at the values in Table~2. The shaded regions indicate parameter ranges that yield realistic discharge rates in the healthy state for all neuronal populations (cf. Sec. \ref{sec:rates}). It is seen that a stronger striato-GPe projection causes larger increases in STN and GPi/SNr rates, and a larger decrease in GPe rate. The change in GPe rate therefore depends on the relative changes in $\nu_{p_2d_2}$, $\nu_{d_1e}$, $\nu_{d_2e}$, $\nu_{ee}$, and $\nu_{ei}$ on the one hand, and $\theta_{p_2}$ and $\theta_{\varsigma}$ on the other, in agreement with the range of experimental findings \citep{Boraud1998, Filion1991, Goldberg2002, Heimer2002, Hutchison1994, Magill2001, Pan1988, Walters2007}. Equilibrium rates for the full parkinsonian state are given in Column $h$ of Table~3. \\ 

\noindent \emph{Effects of remaining projections}\\

We now consider the effects of the various projections that are not included in the classic direct/indirect pathway model. Firing rates are not highly sensitive to local striatal inhibition (cf. Fig. \ref{fig:sensitivity}). Removing it altogether ($\nu_{d_1d_1}=\nu_{d_2d_2}=0$ mV s) slightly increases all rates except those of the GPe and output nuclei, which are decreased (Table~3 Column $i$). The steady-state solutions for $\nu_{d_1d_1}=\nu_{d_2d_2}=0$ mV s in the full parkinsonian state are given in Column $j$ of Table~3. Comparison of the difference between Columns $a$ and $h$, and that between Columns $i$ and $j$, reveals that larger $|\nu_{d_1d_1}|$ and $|\nu_{d_2d_2}|$ help to attenuate alterations in all rates due to dopamine depletion. Increasing $\nu_{d_1s}$ to 0.3 mV s predictably leads to a higher D1 rate, also increasing all other rates except that of the GPi/SNr (Column $k$). Increasing the strength of the thalamo-D2 projection to $\nu_{d_2s} = 0.3$ mV s reduces the GPe rate, which in turn leads to a slightly higher STN rate, elevated GPi/SNr activity, and suppression of corticothalamic targets (Column $l$). To assess the influence of the GPe-GPi/SNr projection we let $\nu_{p_1p_2} = 0$ mV s. This reduces all rates except that of the output nuclei (Column $m$). A stronger STN-GPe projection ($\nu_{p_2\varsigma} = 0.4$ mV s) greatly increases the GPe rate, leading to a reduction in the activity of the output nuclei, which increases corticothalamic and striatal rates (Column $n$). By providing a stronger drive to the STN, a larger $\nu_{\varsigma e}$ increases the rates of both STN and its target nuclei. As a result, thalamic activity is suppressed, reducing both cortical and striatal rates. Column $o$ of Table~3 lists the rates  for $\nu_{\varsigma e}=0.2$ mV s. 

\section{Summary and Discussion}

We have formulated a mean-field model of the basal ganglia-thalamocortical system (BGTCS) that yields realistic steady-state firing rates with parameters in physiologically plausible ranges, and takes into account many projections that were excluded from previous models. Estimates of parameter values and firing rates in health and in Parkinson's disease (PD) were based on an extensive review of the experimental literature. After deriving expressions for the steady states, we have explored the effects of dopamine loss on the average firing rates of the basal ganglia nuclei, thalamus, and cortex. The influence of a range of connection strengths on these changes in firing rates was assessed. The model provides a framework for studying the electrophysiology of the interconnected basal ganglia, thalamus, and cortex, based on previous work by \citet{Rennie1999,Robinson1997,Robinson2001, Robinson2003b, Robinson2005}. Furthermore, it lays the foundation for analysis of dynamics and oscillations in PD, considered in Paper II \citep{vanAlbada2009}. Our main results are the following: 

(i) A decrease in the strength of cortical projections to striatal D1 neurons and a simultaneous increase in that to striatal D2 neurons, as in the direct/indirect pathway model \citep{Albin1989,Alexander1990}, causes elevated GPi/SNr and STN rates and reduced GPe and thalamic rates, in agreement with the findings of many experimental studies of animal models of PD. However, a stronger indirect pathway alone, without modulation of the direct pathway, may be sufficient to approximate empirical results. Therefore, our results are not predicated on a complete separation between the direct and indirect pathways. On the other hand, stronger corticostriatal projections to both D1 and D2 cells, and a simultaneous decrease in striatal firing thresholds, chosen to mimic a reduction in striatal signal-to-noise ratio, do not lead to the expected rate changes. This accords with the findings of \citet{Leblois2006}, but in contrast to that study, we conclude from the experimental literature and our work that dopamine loss often leads to significant changes in firing rates. Note that our results suggest that a combination of reduced corticostriatal connection strengths and striatal firing thresholds of both D1 and D2 populations does not adequately reflect physiological changes upon dopamine loss, but this does not exclude the possibility that dopamine acts as a contrast enhancer \citep{Nicola2004}. This role of dopamine may be mediated by changes in intrinsic properties secondary in importance to changes in synaptic properties \citep{Moyer2007}.

(ii) Besides changes in corticostriatal coupling, several other connection strengths and firing thresholds are likely to be altered in PD.  Loss of mesocortical dopamine is expected to reduce intracortical excitation and especially inhibition \citep{Gao2003, Gulledge2001, Sesack1989,Thurley2008, Zhou1999}. Enhanced enkephalin release may result in weakened lateral inhibition in the GPe \citep{Stanford1999, Terman2002}, while reduced availability of dynorphin is expected to lower the GPe firing threshold \citep{Ogura2000}. Higher levels of extracellular potassium after SNc lesion increase the excitability of the STN \citep{Strauss2008}. In addition, loss of direct dopaminergic innervation stimulates GABA release at striato-GPe terminals, increasing the corresponding connection strength \citep{Floran1997, Querejeta2001}. Modeling results show that reduced intracortical and intrapallidal inhibition and a lower GPe firing threshold help account for the lack of decrease in cortical rate observed in monkeys and rats with nigrostriatal lesions \citep{Dejean2008,Goldberg2002}. 

(iii) Modeling results suggest that changes in corticostriatal coupling strengths are not solely responsible for substantial increases in STN activity observed in animal models of PD \citep{Bergman1994,Kreiss1997, Walters2007}. Reduced intracortical inhibition, a stronger striato-GPe projection, and a lower STN firing threshold all contribute to STN hyperactivity. In the first two cases, the increase in STN activity is associated with a further decrease in GPe rate, whereas a lower STN firing threshold increases both rates. Besides increased excitability, it is also possible that altered inputs from the thalamic parafascicular nucleus \citep{Orieux2000} and/or the PPN \citep{Breit2001, Breit2006, Orieux2000} play a role in STN hyperactivity. The apparent contribution of the PPN to STN hyperactivity found in these studies is paradoxical, because the PPN receives important inhibitory input from the GPi \citep{Yelnik2002}, which is overactive in PD. Furthermore, PPN lesions have been shown to cause symptoms resembling PD in primates \citep{Aziz1998, Kojima1997}. The studies of \citet{Orieux2000} and \citet{Breit2001, Breit2006} were performed in rats, and the results may not generalize to the human situation, since PD is accompanied by significant neuronal loss from the PPN \citep{Pahapill2000, Zweig1989}, which is not observed in the rat model. The STN also receives a direct dopaminergic projection, which probably contributes to changes in STN activity in PD \citep{Blandini2000, Brown1979, Flores1999, Hassani1997, Lavoie1989}. An in vivo study in rats suggested that dopamine inhibits STN firing \citep{Campbell1985}, whereas \citet{Kreiss1997} found that dopamine enhanced STN activity in intact animals, but reduced its activity after SNc lesion. However, a predominantly facilitatory effect has been more frequently reported, both in experiments \citep{Loucif2008,Mintz1986,Ni2001b,Zhu2002}, and in a modeling study \citep{Humphries2006}. Therefore it is unlikely that loss of intrinsic dopamine contributes to increased STN firing in PD.

(iv) Our model provides many approximate bounds on connection strengths in the BGTCS, and shows how the strengths of various projections may account for differences in rates between studies. Parameter estimation is an arduous task because the available data do not cover all relevant aspects of the electrophysiology of the basal ganglia, and there are often inconsistencies between studies. However, our main results are sufficiently robust that they hold for a large part of the physiologically realistic range.

When interpreting the above results, a number of qualifications should be taken into account. First, we have necessarily simplified basal ganglia connectivity, ignoring for instance the projections from GPe to striatum and TRN \citep{Gandia1993, Hazrati1991, Kita1999}, from striatum, STN, and PPN to SNc \citep{Gerfen1992, Jimenez1989, Lavoie1994a}, and from thalamus to STN \citep{Gonzalo2002, Orieux2000}. In addition, the direct dopaminergic innervation from SNc to GPi, GPe, STN, and TRN, which appears to be strongest in GPi \citep{Anaya-Martinez2006, Jan2000, Lavoie1989, Smith1989}, may be incorporated in future studies. Although some of these projections may substantially affect the activity of their targets, their relative influences are difficult to ascertain from the literature. For instance, in addition to its major innervation by the cortex, well-documented reciprocal connections with the thalamic relay nuclei, and inputs from GPe, the TRN receives inputs from diverse brainstem and basal forebrain structures \citep{Cornwall1990}. To keep the model tractable, we have assumed that cortical and thalamic stimuli are the major determinants of TRN activity. Similar arguments hold for the other projections listed above. 

Most of our analysis has also been based on the simplifying assumption that each of the basal ganglia nuclei can be treated as a unit in which neurons have common levels of inputs and outputs. However, studies indicate that each nucleus consists of territories with slightly different connectivity patterns. The division of basal ganglia pathways into sensorimotor, associative, and limbic circuits is the most obvious example of this. In addition, parts of the STN have been reported to project specifically to either GPi or GPe \citep{Gonzalo2002, Parent1989}. Within the GPi, \citet{Parent2001} distinguished between centrally located neurons which terminate in VA, CM-Pf, and PPN, and more peripherally located neurons which innervate the lateral habenula. Moreover, the specific neuronal architectures of the direct and indirect pathways may allow focused excitation of the thalamus via the direct route, and surround inhibition via the indirect route upon cortical stimulation \citep{Haber2004}. Within the framework provided by our model, the influences of additional projections or more detailed connectivity patterns of the basal ganglia can be assessed with relative ease. The dynamics of the present model is explored in detail in Paper II.

\section*{Acknowledgments}
We thank S.C. O'Connor for helpful inputs, and R.T. Gray, C.J. Rennie, P.M. Drysdale, and J.M. Clearwater for useful comments and discussions. This work was supported by the Australian Research Council, an Endeavour International Postgraduate Research Scholarship from the Australian government, and an International Postgraduate Award from the University of Sydney.

\end{document}